\documentclass[10pt]{article}
\usepackage{multicol}
\usepackage{graphicx}
\usepackage{amsmath}
\usepackage[a4paper]{geometry}
\usepackage{enumerate}
\usepackage{hyperref}

\begin{document}

\begin{center}
{\Large \bf Extraction of different temperatures and kinetic freeze-out volume in high energy collisions}

\vskip1.0cm

M.~Waqas$^{1,}${\footnote{Corresponding author. Email (M. Waqas):
waqas\_phy313@yahoo.com, waqas\_phy313@ucas.ac.cn}}, G. X.
Peng$^{1,2}$ {\footnote{Corresponding author. Email (G. X. Peng):
gxpeng@ucas.ac.cn, gxpeng@ihep.ac.cn}}, Muhammad Ajaz; $^{3}$
{\footnote{Corresponding author. Email (M. Ajaz): ajaz@awkum.edu.pk,
muhammad.ajaz@cern.ch}}, Abd Al Karim Haj Ismail $^{4, 5}$,
{\footnote{Email (A.A.K.H.I): a.hajismail@ajman.ac.ae}} Z.~Wazir
$^{6}$, {\footnote{Email (Z.Wazir): zwazir@gudgk.edu.pk}}, Li-Li
Li $^{7}$ {\footnote{Email (Li-Li Li): shanxi-lll@qq.com}},
\\

{\small\it   $^1$ School of Nuclear Science and Technology, University of Chinese Academy of Sciences,
Beijing 100049, People¡'s Republic of China,

$^2$ Theoretical Physics Center for Science Facilities, Institute of High Energy Physics, Beijing 100049, China,

$^3$ Department of Physics, Abdul Wali Khan University Mardan, 23200 Mardan, Pakistan,

$^4$ Department of Mathematics and Science, Ajman University, Ajman P.O. Box 346, United Arab Emirates,

$^5$ Nonlinear Dynamics Research Center (NDRC), Ajman University, Ajman P.O. Box 346, United Arab Emirates,

$^6$ Department of physics, Ghazi University, Dera Ghazi Khan, Pakistan,

$^7$ College of Arts and Sciences, Shanxi Agriculture University, Taigu, Shanxi 030801, China}.

\end{center}

\vskip1.0cm

{\bf Abstract:} We analyze the transverse momentum ($p_T$)
spectra, $1/N_{ev}$[(1/2$\pi$$p_T$) $d^2$$N$/$dyd$$p_T$], of kaon,
proton, deuteron and triton in different centrality events in
gold-gold (Au-Au) collisions at Relativistic Heavy Ion Collisions
(RHIC) by Hagedorn thermal model and extracted the excitation
function of effective temperature, kinetic freeze-out volume,
initial temperature and kinetic freeze-out temperature. We
perceived that the effective temperature, initial temperature and
kinetic freeze-out temperature sharply increases from 7.7 GeV to
14.5 GeV and then remain static from 14.5-39 GeV, and this
consistency may disclose that the onset energy of the phase
transition of partial deconfinement and the whole deconfinement
are 14.5 and 39 GeV respectively. The kinetic freeze-out volume
and mean transverse momentum grows with the rise of collision
energy. Furthermore, the different extracted temperatures are
observed in the order of time evolution of the interacting system,
and they (as well as kinetic freeze-out volume) have an increasing
trend from peripheral to central collisions. We also observed
the mass dependence of the effective temperature and kinetic
freeze-out volume where former increases while the later decreases
for heavier particles, which indicates the early freeze-out of the
heavier particles.
\\



\vskip1.0cm

\begin{multicols}{2}

{\section{Introduction}} Temperature is surely one of the most
prominent concept in physics due to its wide application in
experiments and theoretical analysis. At least four kinds of
temperatures are frequently used in the physics of high energy
heavy ion collisions which include the initial temperature
($T_i$), chemical freeze-out temperature ($T_{ch}$), kinetic
freeze-out temperature or thermal freeze-out temperature ($T_0$)
and effective temperature. These temperatures are correspondent to
different collision stages. The excitation degree of interacting
system at initial stage of collision is described by the initial
temperature and it is the  preliminary temperature where hadrons
interact both elastically and in-elastically in the hadronic
medium. The initial temperature has less studies in the community
due to undefined method, though it should be based on the
transverse momentum spectra. The initial temperature is followed
by the chemical freeze-out temperature which occurs at the stage
of chemical freeze-out, when the inelastic collisions cease, where
it describes the excitation degree of interacting system at the
chemical freeze-out stage. After this stage, no further new
particles are produced and the yield of particles become fixed
[1--4]. Various thermodynamic models can be used to get the
information of this stage in terms of chemical freeze-out
temperature and baryon chemical potential [5--8]. Subsequently,
the particles experience the elastic collisions only. The
inter-particle dissociation grows after further expansion of the
system, where the elastic collisions between the particles cease
which leads to the kinetic freeze-out stage and there is no longer
change in the $p_T$ spectra after this stage. Various
hydrodynamical models can be used be used to extract the kinetic
freeze-out properties [5, 6, 9, 10, 11, 12, 13]. The kinetic
freeze-out temperature $(T_0)$ and radial flow velocity $(\beta)$
characterizes the kinetic freeze-out stage, and it carry the
signatures of the transverse expansion of the system. The kinetic
freeze-out temperature describes the excitation of interacting
system at the kinetic freeze-out stage, and it can be obtained
from the transverse momentum spectra of the particles by getting
rid of the transverse flow effect and leaving only the
contribution of random thermal motion [7, 10, 14--16]. In case,
the temperature extracted from $p_T$ spectra also contains the
contribution of transverse flow, this type of temperature is
called effective temperature ($T$) [17, 18], which is not a real
temperature, but can be obtained from the spectra of heat
distribution. The effective temperature occurs at the same time or
a little before than kinetic freeze-out temperature. Besides,
volume is one of the important parameters to be studied. Different
stages correspond to different freeze-out volumes but we are
interested in kinetic freeze-out volume. The volume occupied by
the hadrons when they decouple from the system is called kinetic
freeze-out volume, and it is very important for the extraction of
multiplicity, micro-canonical heat capacity and its negative
branch or shape of the caloric curves under the thermal
constraints [19--23].

Lattice quantum chromodynamics (QCD) is very important in the
study of strong interactions and it predicts the transition
between the hadronic matter and quark-gluon plasma (QGP) to be a
cross-over [24, 25] at $\mu_B$=0. Several QCD based models
[26--29] and calculations [26] suggest  that a system created in
collisions corresponding to larger $\mu_B$, the phase transition
is the first order. In $T_{ch}$-$\mu_B$ plane , the point where
the first order phase transition ends, is referred as the critical
end point (CEP) of QCD [30, 31]. Besides, let consider a certain
volume containing baryons. It is quite understandable that baryons
have non-zero spatial volume [32]. This clearly indicates that a
critical volume exists where the baryons fill the volume
completely and it is assumed that the baryon structure vanishes at
the critical volume and forms the quark gluon plasma.

In heavy ion collisions, the production mechanism of light nuclei
is a subject of intense debate. In fact, low-energy heavy-ion
physics has clearly more connection with nuclear structure
physics, for example the studied heavy ions can split up into
smaller nuclei that can be scrutinized. Whereas, the collided ions
in ultra-relativistic heavy-ion collisions are mainly creating a
region of large energy density and temperature that is nearly free
of baryon number. e.g. the baryon chemical potential ($\mu_B$) at
low energies is about 1 GeV and is close to zero at LHC. This
means that the anti-protons and anti-neutrons are equally produced
at LHC as their matter counter pieces. $\mu_B$ is still some MeV
at RHIC, which leads to slight variation in the production of
protons and anti-protons that results in a small variation of the
production of nuclei and anti-nuclei.

In relativistic heavy ion collision, the fundamental mechanism for light nuclei production is not well understood.
The light nuclei have small binding energy, therefore they can not persist when the temperature becomes much higher
than their binding energy. For light hadrons, the typical kinetic freeze-out temperature is approximately 100 MeV,
therefore they may disintegrate and be formed again by coalescence after the decoupling of nucleons.

In this paper we shall study the $p_T$ spectra of kaon, proton,
deuteron and triton in different centrality intervals in Au-Au
collisions at various energies at Relativistic Heavy Ion Collision
(RHIC). We used the Hagedorn model to extract the effective
temperature, initial temperature, kinetic freeze-out temperature,
mean transverse momentum, kinetic freeze-out volume by fitting the
experimental data of $p_T$ spectra. Among the above parameters,
initial temperature and kinetic freeze-out temperature are
extracted by an indirect method which will be discussed in section
3.
\\

{\section{The method and formalism}} There are two general
processes for the particles production. (1) soft excitation
process (2) hard scattering process. Soft process contributes in a
narrow $p_T$ range of 0$\sim$2 or 3 GeV/c and is responsible for
the production of light flavored particles. Soft excitation has
various choices e.g. Hagedorn model [33, 34], blast wave model
with Boltzmann Gibbs statistics [10, 35--37], blast wave model
with Tsallis statistics [38--41], Tsallis distribution [42--44]
and Standard distribution [45, 46]. Although some particles are
produced at not too high energies and are distributed in a wide
$p_T$ range but the contribution of hard scattering can be
neglected and the soft excitation is the main contributor for the
production of particles. There are limited choices for the hard
scattering and can be described by the theory of strong
interaction [47--49]. The contribution of hard scattering process
is parameterised to an inverse power law [49, 50], i:e Hagedorn
function [33, 51]
\begin{align}
f(p_T)=\frac{1}{N}\frac{dN}{dp_T}= Ap_T \bigg( 1+\frac{p_T}{
p_0} \bigg)^{-n},
\end{align}
where A is the normalization constant, and $p_0$
and n are the free parameters.

According to [33], the hagedorn model in terms of probability density function at mid-rapidity
results in the transverse momentum ($p_T$) spectra distribution as
\begin{align}
f(p_T)=\frac{1}{N}\frac{\mathrm{d}N}{\mathrm{d}p_\mathrm{T}}=\frac{gV}{(2\pi)^2}  p_T{\sqrt{p_T^2+m_0^2}}\nonumber\\
\sum\limits_{n=1}^{\infty}(S)^{n+1}K_1\bigg(n\frac{\sqrt{p_T^2+m_0^2}}{T}\bigg),
\end{align}
where  N is the number of particles, g is the degeneracy factor,
$V$ is kinetic freeze outvolume, $m_0$ is the rest mass of the
particle and K is the modified Bessel function of second kind.
S can be +1 or -1. S=+1 is the Fermi system and S=-1 is the
Bose system, and $n$ is the Hagedorn index which is different for
different particles, i.e $n$ can be $1-5$ for kaon, 1 for proton,
and for deuteron and triton it is $1-2$. $n$ has no connection
with the Fermi or Bose system.

Considering the experimental rapidity range [$y_{min}$, $y_{max}$] around the mid-rapidity, the accurate form of eq.(2) becomes
\begin{align}
f(p_T)=\frac{1}{N}\frac{\mathrm{d}N}{\mathrm{d}p_\mathrm{T}}=\frac{gV}{(2\pi)^2}  p_T{\sqrt{p_T^2+m_0^2}}\nonumber\\
\int_{y_{\min}}^{y_{\max}}coshy\sum\limits_{n=1}^{\infty}(S)^{n+1}K_1\bigg(n\frac{\sqrt{p_T^2+m_0^2}coshy}{T}\bigg)dy,
\end{align}
Eq. (3) maybe different from Eq.(2). Eq. (3) is more appropriate to the particles produced in a wide $p_T$ range
compared to eq. (2). We can change the rapidity range by adding or subtracting a rapidity shift to cover the mid-rapidity
 if the rapidity range does not cover the mid-rapidity and thus the contribution of directional movement of the emission source
 can be excluded.

The single component of Hagedorn model is not enough for the simultaneous description of low $p_T$ region, so we have used
the two-component Hagedorn model in recent work. The Hagedorn model with multi-component can be expressed as

\begin{align}
f(p_T)=\frac{1}{N}\frac{\mathrm{d}N}{\mathrm{d}p_\mathrm{T}}=\sum\limits_{i=1}^{l}k_i\frac{gV_i}{(2\pi)^2}p_T\nonumber\\
{\sqrt{p_T^2+m_0^2}}\sum\limits_{n=1}^{\infty}(S)^{n+1}K_1\bigg(n\frac{\sqrt{p_T^2+m_0^2}}{T_i}\bigg)
\end{align}
or
\begin{align}
f(p_T)=\frac{1}{N}\frac{\mathrm{d}N}{\mathrm{d}p_\mathrm{T}}= \sum\limits_{i=1}^{l}k_i\frac{gV_i}{(2\pi)^2} p_T{\sqrt{p_T^2+m_0^2}}\nonumber\\
\int_{y_{\min}}^{y_{\max}}coshy\sum\limits_{n=1}^{\infty}(S)^{n+1}K_1\bigg(n\frac{\sqrt{p_T^2+m_0^2}coshy}{T_i}\bigg)dy,
\end{align}
where l represents the number of components and $k_i$ denotes
contribution fraction, $T_i$ is the is effective temperature and
$V_i$ is the kinetic freeze out volume corresponding to i-th
component. In case of multi-components, we have used the
superposition principle. We would like to clarify that Eq.
(2)-(5) are based on both Eq. (2.2) and (B.6) of ref. [33].
\begin{align}
f(p_T)=\frac{1}{N}\frac{dN}{dp_T}=kf_S(p_T)+(1-k)f_H(p_T),
\end{align}
where $f_S$ is the soft process with the contribution fraction of $k$, and $f_H$ is the hard components with $(1-k)$ contribution fraction.
In eq.(5), the soft process contributes from 0$\sim$2 or 3 GeV/c and hard scattering contributes the whole $p_T$ region.

We may also use the usual step function in case of two component Hagedorn model
\begin{align}
f_0(p_T)=A_1\theta(p_1-p_T) f_S(p_T) + A_2 \theta(p_T-p_1)f_H(p_T),
\end{align}
where $A_1$ and $A_2$ are the normalization constants which
synthesize $A_1$$f_S$$(p_1)$=$A_1$$f_H$$(p_1)$ and $\theta$(x) is
the usual step function. It should be noted that $f_S$ and
$f_H$ are just the two components but not the different identified
particles. $T_1$ and $T_2$ in table 1, 2, 3 and 4 are the soft and
hard components respectively.

 At the end, we would like to point out that the Hagedron thermal model is usually
associated with continuous mass spectra but we have used the $p_T$ spectra in the present work.
Actually both the $p_T$ and $m_T$ spectra can be inter-convertable, i.e , in order to transform the probability density function
$f_S(p_T)$ of $p_T$ to the probability density function $f_{S'}(m_T)$ of $m_T$, we have a relation
\begin{align}
f_S(p_T)|dp_T|=f_{S'}(m_T)|dm_T|,
\end{align}
Then
\begin{align}
f_{S'}(m_T)|dm_T|=\frac{m_T}{\sqrt{{m^2_T}-{m^2_0}}}f_{p_T}({\sqrt{{m^2_T}-{m^2_0}}}),
\end{align}
\\

{\section{Results and discussion}} Figure 1 and 2 demonstrate
the event centrality dependent double differential transverse
momentum ($p_T$) spectra, [(1/2$\pi$$p_T$) $d^2$$N$/$dyd$$p_T$] of
$K^+$ and $p$ respectively, produced in Au-Au collisions at
RHIC-BES. The spectra of
\begin{figure*}[htbp]
\begin{center}
\includegraphics[width=14.cm]{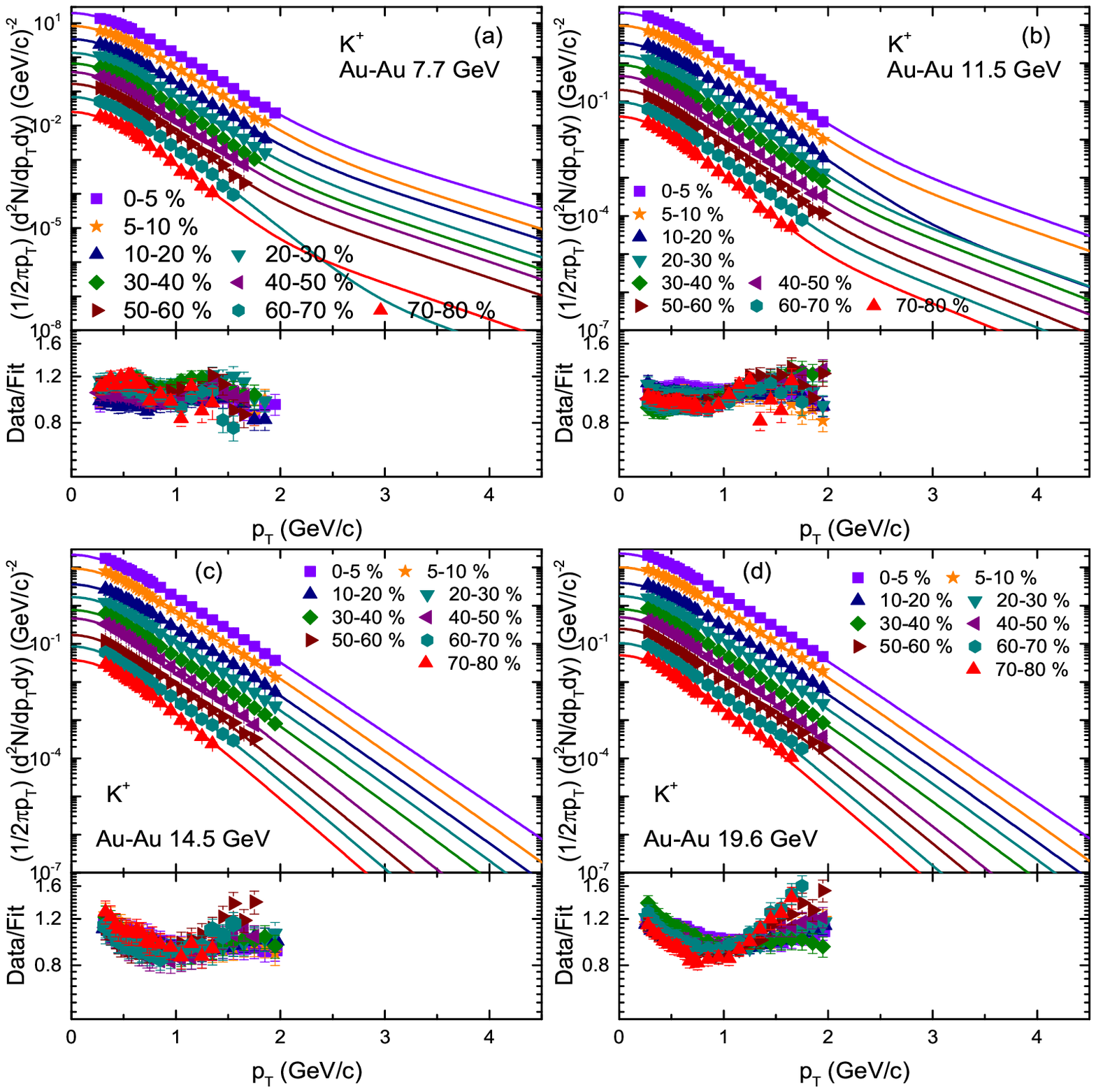}
\end{center}
continue
\end{figure*}
\begin{figure*}[htbp]
\begin{center}
\includegraphics[width=14.cm]{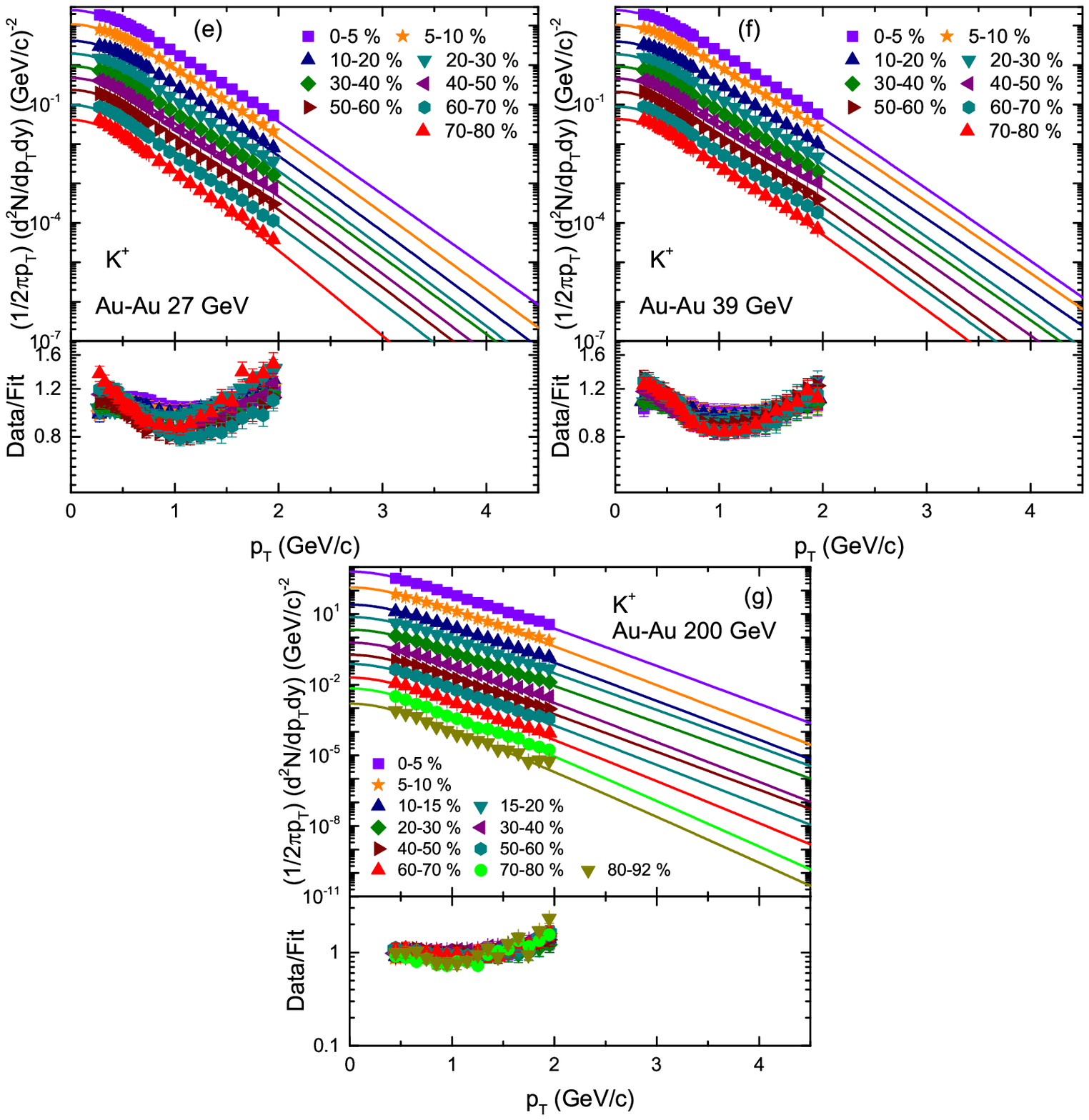}
\end{center}
Fig. 1. Transverse momentum spectra of $K^+$ produced in different
centrality intervals in Au-Au collisions. The symbols represent
the experimental data measured by STAR Collaboration at RHIC [52,
53], while the curves are our fitting results by using the
Hagedorn thermal model. The corresponding ratio of data/fit is
presented in each panel. The spectra of $K^+$ are scaled by
in different centrality intervals in order to present the figure
clearly. The spectra of $K^+$ at 7.7-39 GeV in 5--10\%, 10--20\%,
20--30\%, 30--40\%, 40--50\%, 50--60\%, 60--70\% and 70--80\%
centrality bins are scaled by the factor 1/2, 1/4, 1/6, 1/8, 1/10,
1/12, 1/14 and 1/16 respectively, however at 200 GeV the spectra
in 5--10\%, 10--15\%, 15--20\%, 20--30\%, 30--40\%, 40--50\%,
50--60\%, 60--70\%, 70--80\% and 80--92\% centrality bins are
scaled by the factor 1/2, 1/4, 1/6, 1/9, 1/14, 1/25, 1/40, 1/70,
1/120 and 1/230 respectively.
\end{figure*}
\begin{figure*}[htbp]
\begin{center}
\includegraphics[width=14.cm]{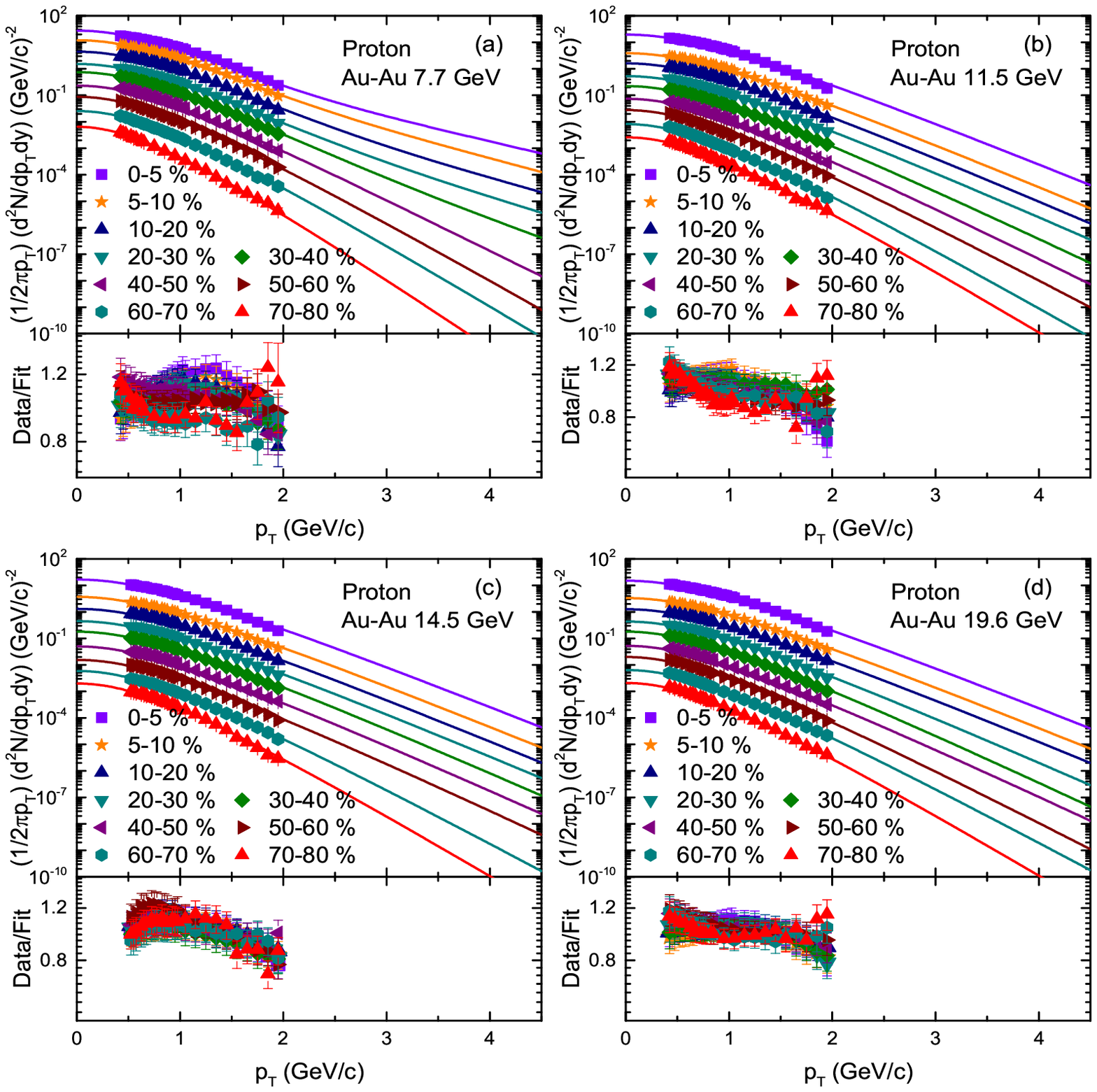}
\end{center}
continue
\end{figure*}
\begin{figure*}[htbp]
\begin{center}
\includegraphics[width=14.cm]{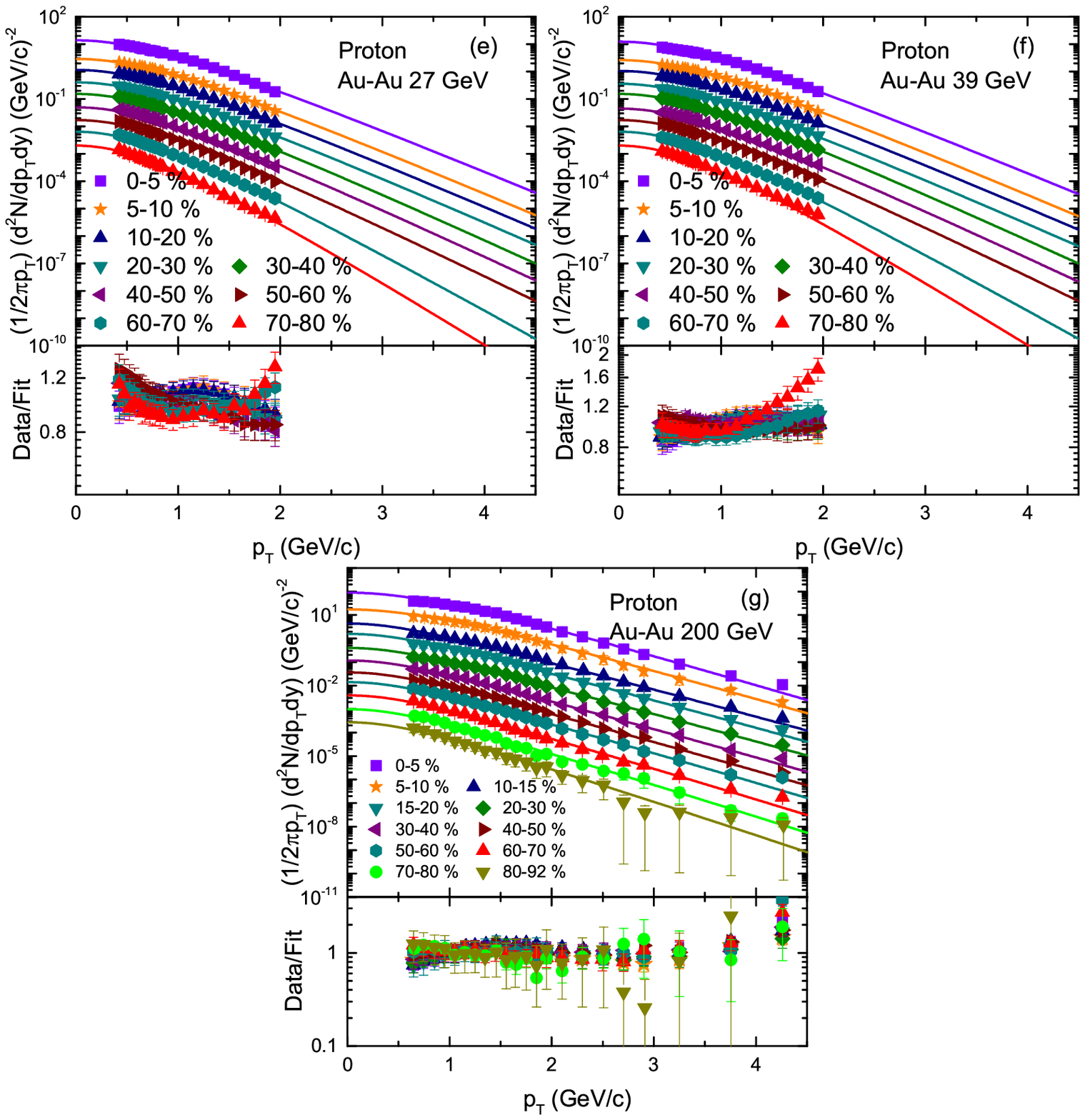}
\end{center}
Fig. 2. Transverse momentum spectra of proton produced in
different centralities in Au-Au collisions. The symbols represent
the experimental data measured by STAR Collaboration at RHIC [52,
53], while the curves are our fitting results by using the
Hagedorn thermal model. The corresponding ratio of data/fit is
presented in each panel. The spectra of $p$ at 7.7 GeV in
5--10\%, 10--20\%, 20--30\%, 30--40\%, 40--50\%, 50--60\%,
60--70\% and 70--80\% centrality bins are scaled by the factor
1/2, 1/4, 1/8, 1/12, 1/22, 1/40, 1/80 and 1/180 respectively, and
the spectra at 11.5-39 GeV in 5--10\%, 10--20\%, 20--30\%,
30--40\%, 40--50\%, 50--60\%, 60--70\% and 70--80\% centrality
bins are scaled by the factor 1/4, 1/8, 1/16, 1/28, 1/55, 1/90,
1/160 and 1/270, however at 200 GeV the spectra in 5--10\%,
10--15\%, 15--20\%, 20--30\%, 30--40\%, 40--50\%, 50--60\%,
60--70\%, 70--80\% and 80--92\% centrality bins are scaled by the
factor 1/2, 1/4, 1/6, 1/9, 1/14, 1/25, 1/40, 1/70, 1/120 and 1/230
respectively.
\end{figure*}
$K^+$ and $p$ in fig. 1 and 2 respectively from panel (a)-(f)
are distributed in to 0--5\%, 5--10\%, 10--20\%, 20--30\%,
30--40\%, 40--50\%, 50--60\%, 60--70\% and 70--80\% centrality
bins, while in panel (g) their spectra are distributed in to
0--5\%, 5--10\%, 10--15\%, 15--20\%, 20--30\%, 30--40\%, 40--50\%,
50--60\%, 60--70\%, 70--80\% and 80--92\% centrality bins. The
symbols are the experimental data measured by the STAR
Collaboration at $|y|<0.1$ [52] in panels (a)-(f) in fig. 1 and 2,
and in $|\eta|<0.5$ [53] in panel (g) in fig. 1 and 2, and the
curves are our fitting results by the Hagedorn thermal model. Each
panel is followed by the result of its data/fit ratios to monitor
the difference between the data and fit. The results of related
parameters along with $\chi^2$ and degree of freedom (dof) for
$K^+$ and $p$ respectively, are listed in Table 1 and 2. One can
see that the $p_T$ spectra of $K^+$ and $p$ are shown to obey
approximately the Hagedorn thermal model.

Figure 3 and 4 are similar to fig.1 and 2, but they present the
$p_T$ spectra of $d$ and $t$ produced in Au-Au collisions. The
spectra of $d$ are distributed in 0--10\%, 10--20\%, 20--40\%,
40--60\% and 60--80\% centrality bins in fig. 3, while the spectra
of $t$ are distributed in 0--10\%, 10--20\%, 20--40\% and 40--80\%
centrality classes in fig. 4. The symbols are cited from the
experimental data measured by the STAR Collaboration at
mid-rapidity $|y|<0.3$ [54] in fig. 3 and $|y|<0.5$ [55] in fig.
4, and the curves are our fitting results by the Hagedorn thermal
model. The ratios of data/fit are correspondingly presented in the
lower case of each panel. The results of related parameters along
with $\chi^2$ and degree of freedom (dof) for deuteron and triton
respectively, are listed in Table 3 and 4. One can see that the
$p_T$ spectra of deuteron and triton are shown to obey
approximately the Hagedorn thermal model.

\begin{figure*}[htbp]
\begin{center}
\includegraphics[width=14.cm]{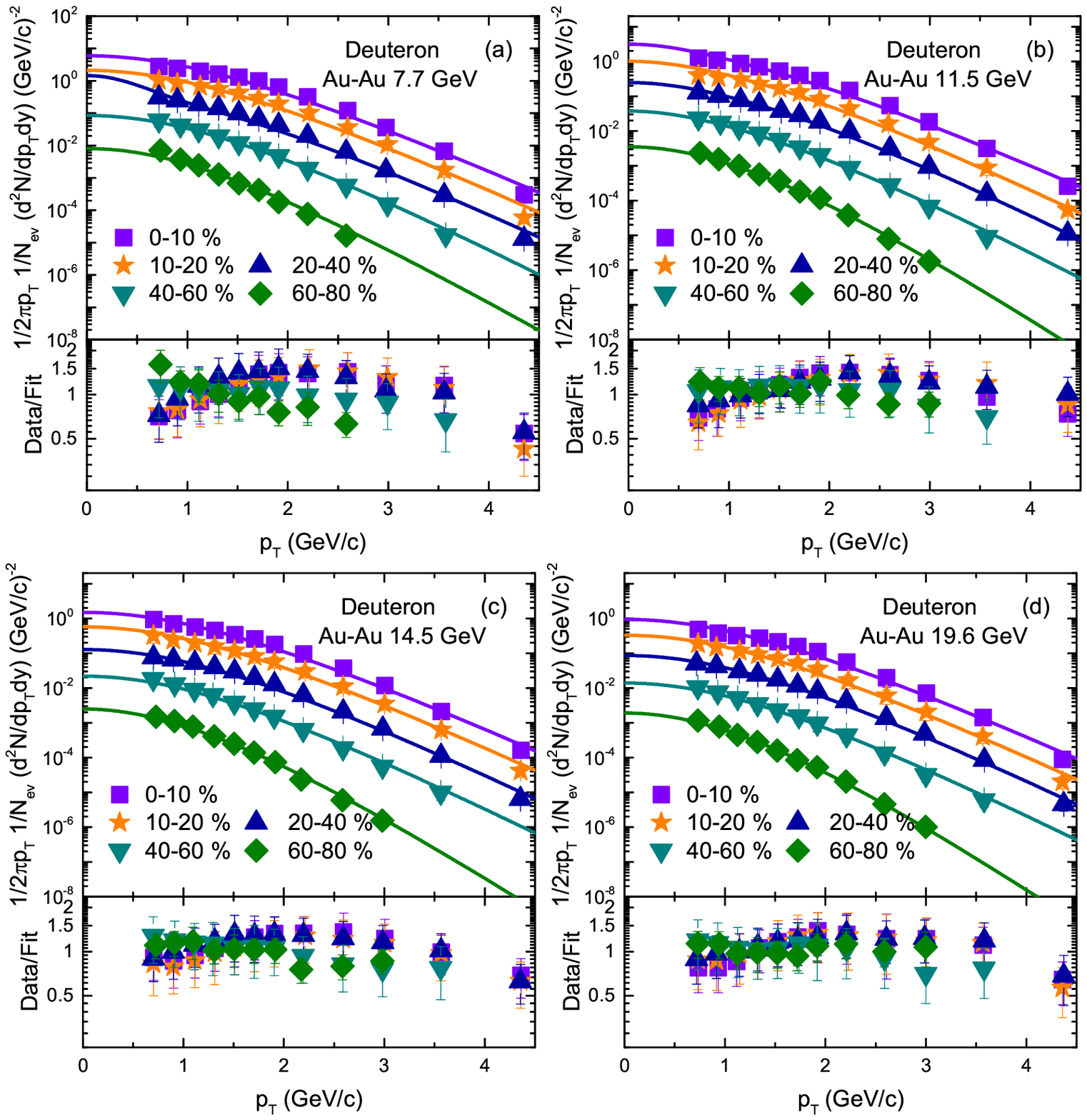}
\end{center}
continue
\end{figure*}
\begin{figure*}[htbp]
\begin{center}
\includegraphics[width=14.cm]{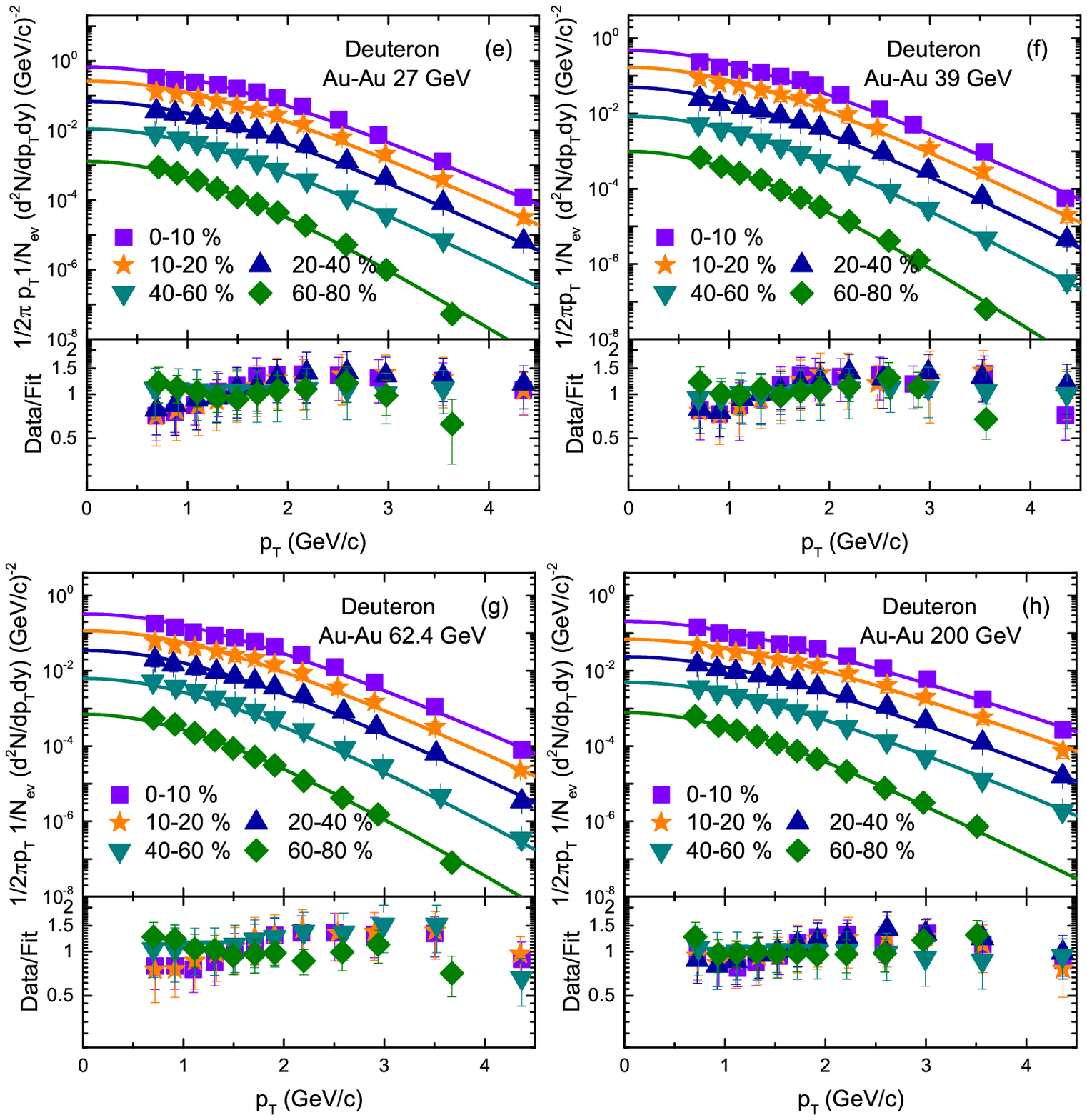}
\end{center}
Fig. 3. Transverse momentum spectra of deuteron produced in
different centralities in Au-Au collisions at
 mid-rapidity $|y|<0.3$ [54]. The symbols represent
the experimental data measured by STAR Collaboration at RHIC,
while the curves are our fitted results by using the Hagedorn
thermal model. The corresponding ratio of data/fit is presented in
each panel.
\end{figure*}

\begin{figure*}[htbp]
\begin{center}
\includegraphics[width=14.cm]{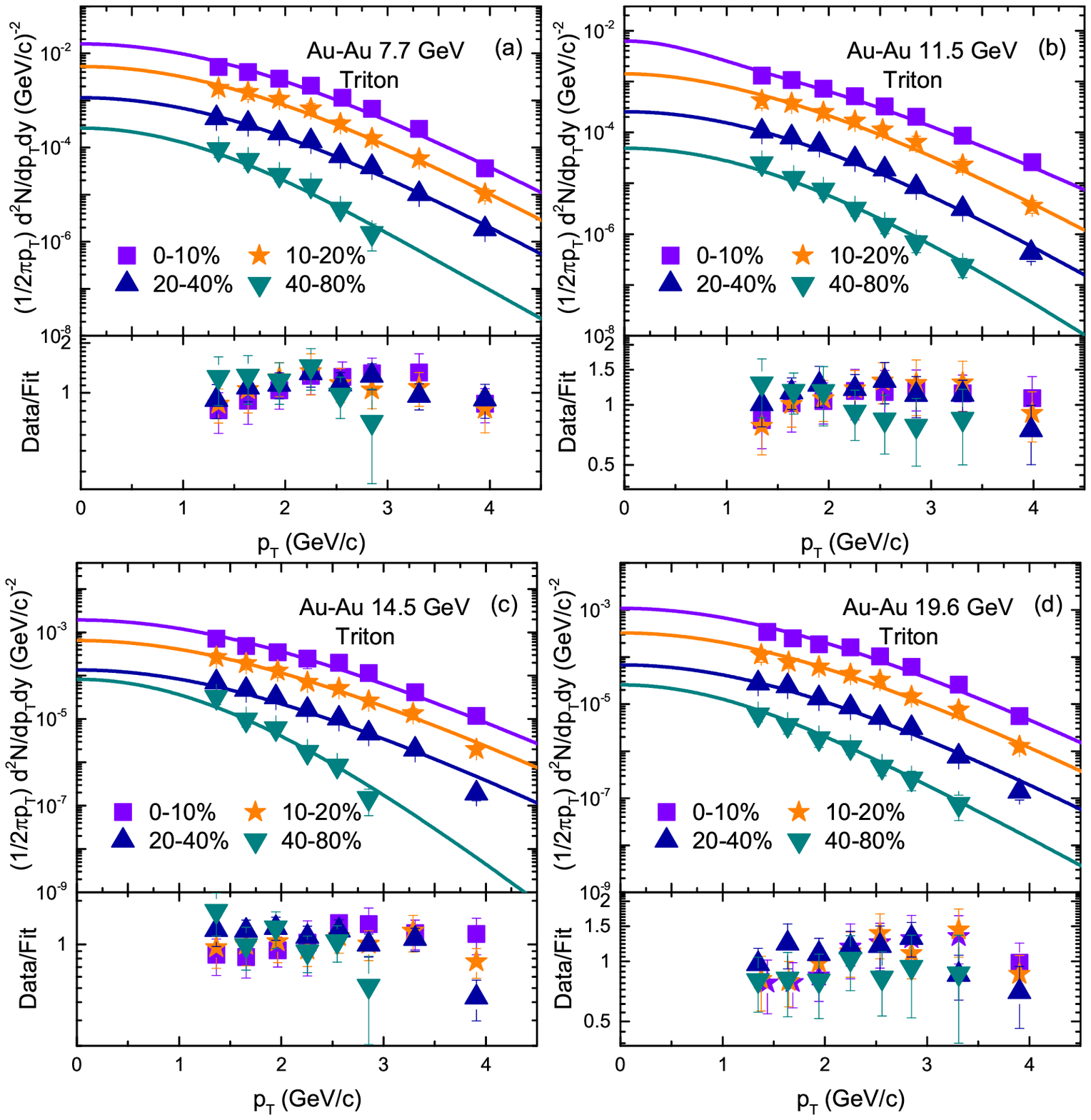}
\end{center}
continue
\end{figure*}
\begin{figure*}[htbp]
\begin{center}
\includegraphics[width=14.cm]{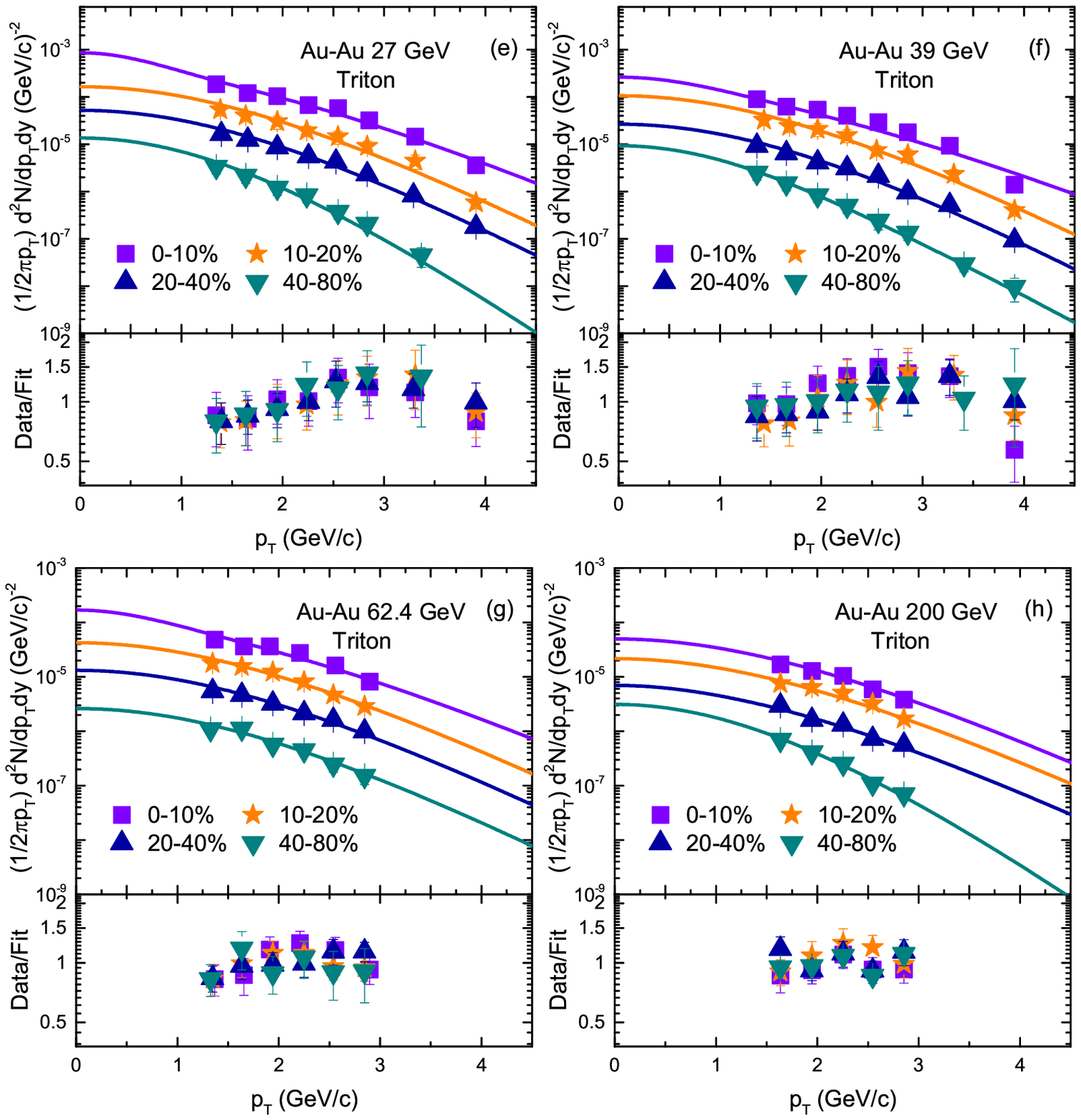}
\end{center}
Fig. 4. Transverse momentum spectra of triton produced in
different centralities in Au-Au collisions at
 mid-rapidity $|y|<0.3$ [55]. The symbols represent
the experimental data measured by STAR Collaboration at RHIC,
while the curves are our fitted results by using the Hagedorn
thermal model. The corresponding ratio of data/fit is presented in
each panel.
\end{figure*}

\begin{figure*}[htbp]
\begin{center}
\includegraphics[width=14.cm]{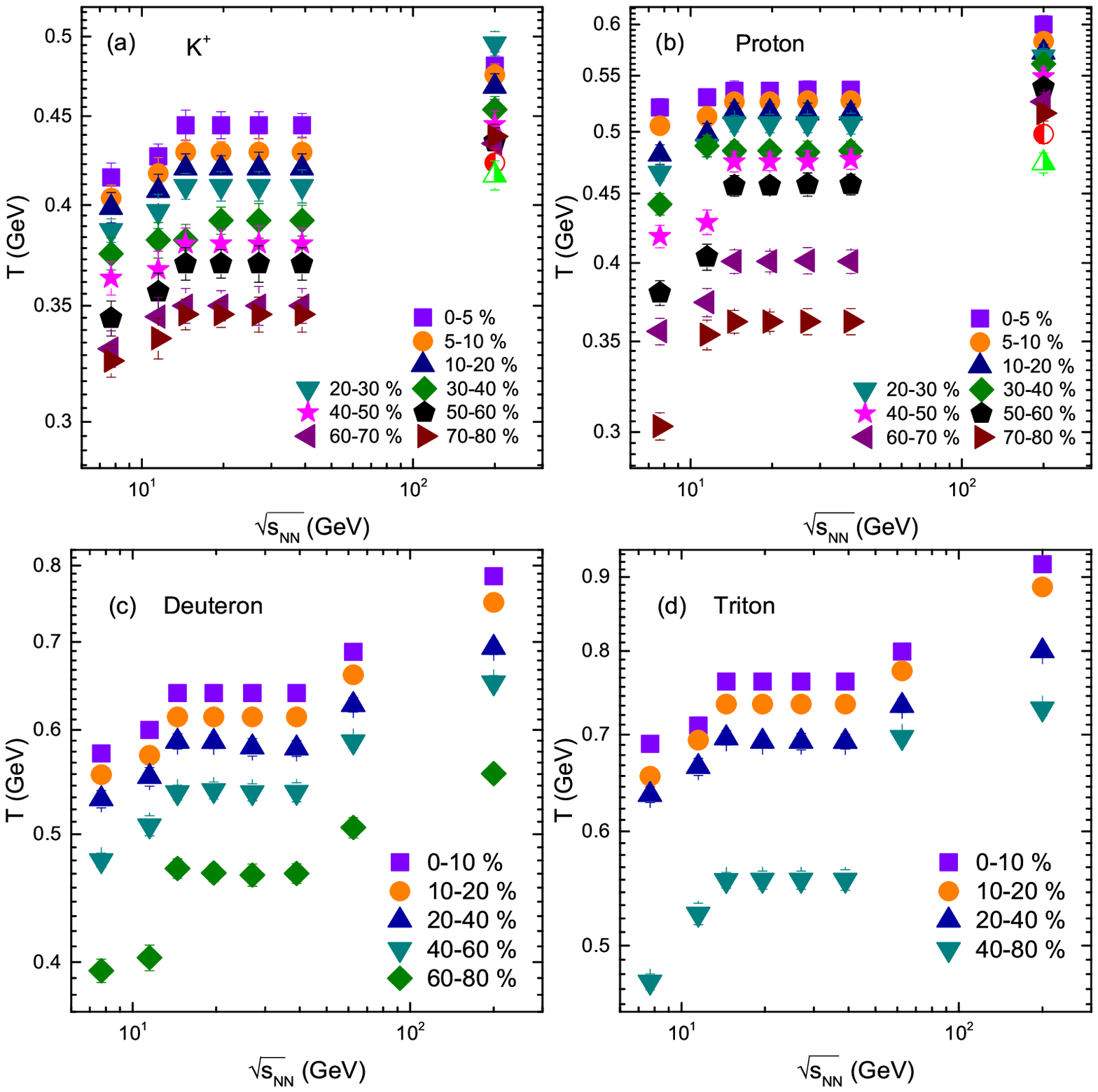}
\end{center}
Fig. 5. Dependence of effective temperature on energy and
centrality for $K^+$, $p$, $d$ and $t$. The centrality bins
at the center of mass energy from 7.7-39 GeV are same and they
shown in each panel, while the centrality bins at 200 GeV are
different and are not presented because the figure will be much
crowded.
\end{figure*}
To study the changing trends of parameters, figure 5 shows the
dependence of effective temperature on energy and centrality.
Panel (a)-(d) show the results for $K^+$, $p$, $d$ and $t$
respectively. Different symbols with different colors represent
different centrality bins. The legends presented inside the
figure are for the energy range of 7.7-39 GeV, and that of 200 GeV
are not included in the figure due to the reason that its
centrality intervals are different from the others. The centrality
intervals of 200 GeV are given above in fig.1 and 2. The trend of
symbols from up to downward shows the dependence of effective
temperature on centrality and from left to right shows its
dependence on energy. We noticed a significant increase of the
effective temperature from 7.7 to 14.5 GeV, and then the trend
becomes consistent from 14.5 to 39 GeV, which later on was again
increased at maximum energy. We will discuss the behavior of
excitation function of $T$ as that of $T_0$ later. In addition,
the effective temperature is smaller in peripheral collisions and
it increases towards central collisions. It is also observed
that the effective temperature is increasing with increasing the
particles mass, which exhibits a scenario of multiple freeze-out.
Larger $T$ for heavier particles indicate their early freeze-out.
One can see that $t$ ($K^+$) has the largest (lowest) values of
$T$ among the studied particles which means that $t$ ($K^+$)
freeze-out early (later).

\begin{table*}
{\scriptsize Table 1. Values of free parameters ($T_1$, $T_2$£¬
$V_1$, $V_2$ and $k$), normalization constant ($N_0$), $\chi^2$,
and degree of freedom (dof) corresponding to the curves in Fig. 1.
\vspace{-.50cm}
\begin{center}
\begin{tabular}{ccccccccccc}\\ \hline\hline
Collisions       & Centrality   & $T_1$ (GeV)     & $T_2$ (GeV)
& $V_1 (fm^3)$ & $V_2 (fm^3)$   & $k$              & $N_0$
& $\chi^2$/ dof \\ \hline
Fig. 1           & 0--5\%       &$0.414\pm0.009$ & $0.895\pm0.008$  & $3600\pm210$ & $2430\pm190$   & $0.97\pm0.19$    &$1.2\pm0.04$         & 4/22\\
Au-Au            & 5--10\%      &$0.390\pm0.008$  & $0.845\pm0.008$  & $3500\pm170$ & $2350\pm200$   & $0.97\pm0.18$    &$0.5\pm0.03$         & 7/22\\
7.7 GeV          & 10--20\%     &$0.380\pm0.007$  & $0.845\pm0.008$  & $3500\pm160$  & $2150\pm100$   & $0.96\pm0.20$    &$0.20\pm0.03$        & 17/22\\
                 & 20--30\%     &$0.367\pm0.008$  & $0.805\pm0.008$  & $3300\pm168$ & $2136\pm188$   & $0.955\pm0.16$   &$0.08\pm0.004$       & 10/22\\
                 & 30--40\%     &$0.357\pm0.006$  & $0.805\pm0.007$  & $3100\pm174$ & $2136\pm140$   & $0.96\pm0.20$    &$0.04\pm0.004$       & 16/21\\
                 & 40--50\%     &$0.345\pm0.007$  & $0.802\pm0.007$  & $3000\pm171$ & $2036\pm160$   & $0.96\pm0.14$    &$0.022\pm0.004$      & 7/20\\
                 & 50--60\%     &$0.330\pm0.008$  & $0.800\pm0.008$  & $2800\pm149$ & $2036\pm140$   & $0.97\pm0.16$    &$0.01\pm0.003$       & 13/19\\
                 & 60--70\%     &$0.330\pm0.009$  & $0.800\pm0.007$  & $2808\pm156$ & $1823\pm190$   & $0.999\pm0.19$   &$0.004\pm0.0004$     & 10/18\\
                 & 70--80\%     &$0.316\pm0.008$  & $0.780\pm0.008$  & $2808\pm170$ & $1623\pm154$   & $0.98\pm0.18$    &$0.00144\pm0.0003$   & 8/15\\
\cline{2-8}
Fig. 1           & 0--5\%       &$0.413\pm0.007$ & $0.869\pm0.006$  & $3900\pm208$  & $2230\pm170$   & $0.97\pm0.12$    &$6.25\pm0.6$         & 12/22\\
Au-Au            & 5--10\%      &$0.403\pm0.006$  & $0.869\pm0.007$  & $3900\pm190$ & $2230\pm160$   & $0.97\pm0.16$    &$0.55\pm0.05$        & 4/23\\
11.5 GeV         & 10--20\%    &$0.403\pm0.007$  & $0.861\pm0.006$  & $3700\pm190$ & $2230\pm160$   & $0.99\pm0.17$    &$0.20\pm0.05$        & 6/23\\
                 & 20--30\%     &$0.383\pm0.009$  & $0.840\pm0.006$  & $3600\pm181$ & $2100\pm181$   & $0.97\pm0.15$    &$0.09\pm0.003$       & 8/23\\
                 & 30--40\%     &$0.363\pm0.009$  & $0.780\pm0.005$  & $3500\pm193$ & $2000\pm170$   & $0.955\pm0.21$   &$0.05\pm0.003$       & 18/23\\
                 & 40--50\%     &$0.350\pm0.006$  & $0.780\pm0.008$  & $3400\pm169$ & $1900\pm176$   & $0.96\pm0.10$    &$0.025\pm0.005$      & 12/23\\
                 & 50--60\%     &$0.340\pm0.006$  & $0.780\pm0.005$  & $3200\pm163$ & $1900\pm163$   & $0.96\pm0.14$    &$0.011\pm0.003$      & 11/23\\
                 & 60--70\%     &$0.328\pm0.008$  & $0.750\pm0.006$  & $3200\pm169$ & $1900\pm170$   & $0.96\pm0.11$    &$0.005\pm0.0004$     & 15/23\\
                 & 70--80\%     &$0.318\pm0.006$  & $0.745\pm0.007$  & $3551\pm150$ & $1700\pm103$   & $0.96\pm0.10$    &$0.002\pm0.0004$     & 6/23\\
\cline{2-8}
Fig. 1           & 0--5\%       &$0.429\pm0.009$  & $0.469\pm0.007$  & $4200\pm209$ & $2430\pm140$  & $0.61\pm0.11$    &$1.5\pm0.2$          & 3/23\\
Au-Au            & 5--10\%      &$0.434\pm0.008$  & $0.420\pm0.006$  & $4200\pm190$ & $2200\pm110$   & $0.62\pm0.13$    &$0.65\pm0.05$        & 4/23\\
14.5 GeV         & 10--20\%     &$0.429\pm0.009$  & $0.420\pm0.008$  & $4200\pm150$ & $2000\pm120$   & $0.61\pm0.08$    &$0.25\pm0.04$        & 3/23\\
                 & 20--30\%     &$0.420\pm0.006$  & $0.420\pm0.007$  & $4000\pm119$ & $2000\pm103$   & $0.60\pm0.05$    &$0.12\pm0.03$        & 18/23\\
                 & 30--40\%     &$0.410\pm0.007$  & $0.410\pm0.006$  & $3800\pm160$ & $2000\pm110$   & $0.62\pm0.07$    &$0.07\pm0.004$       & 16/23\\
                 & 40--50\%     &$0.370\pm0.008$  & $0.400\pm0.007$  & $3800\pm162$ & $1800\pm123$   & $0.62\pm0.08$    &$0.03\pm0.003$       & 11/23\\
                 & 50--60\%     &$0.370\pm0.008$  & $0.370\pm0.007$  & $3600\pm160$ & $1620\pm103$   & $0.62\pm0.04$    &$0.011\pm0.003$      & 10/23\\
                 & 60--70\%     &$0.360\pm0.006$  & $0.350\pm0.008$  & $3600\pm166$ & $1620\pm100$   & $0.62\pm0.14$    &$0.005\pm0.0004$     & 6/23\\
                 & 70--80\%     &$0.350\pm0.006$  & $0.346\pm0.006$  & $3400\pm150$ & $1620\pm103$   & $0.62\pm0.10$   & $0.0024\pm0.0004$   & 9/23\\
\cline{2-8}
Fig. 1           & 0--5\%       &$0.429\pm0.007$  & $0.469\pm0.006$  & $4600\pm220$ & $2430\pm150$   & $0.61\pm0.11$     &$1.5\pm0.2$          & 14/23\\
Au-Au            & 5--10\%      &$0.434\pm0.005$  & $0.420\pm0.005$  & $4300\pm200$ & $2430\pm119$   & $0.62\pm0.10$     &$0.65\pm0.05$        & 9/23\\
19.6 GeV         & 10--20\%     &$0.429\pm0.006$  & $0.420\pm0.008$  & $4640\pm178$ & $2430\pm100$   & $0.61\pm0.09$    &$0.25\pm0.04$         & 13/23\\
                 & 20--30\%     &$0.420\pm0.006$  & $0.420\pm0.008$  & $4200\pm175$ & $2200\pm118$   & $0.60\pm0.08$    &$0.12\pm0.03$         & 5/23\\
                 & 30--40\%     &$0.410\pm0.006$  & $0.410\pm0.006$  & $4000\pm180$ & $2225\pm132$   & $0.62\pm0.07$    &$0.07\pm0.004$        & 16/23\\
                 & 40--50\%     &$0.370\pm0.009$  & $0.400\pm0.006$  & $3930\pm162$ & $2100\pm133$   & $0.62\pm0.08$    &$0.03\pm0.003$        & 12/23\\
                 & 50--60\%     &$0.370\pm0.008$  & $0.370\pm0.006$  & $4000\pm150$ & $1800\pm116$   & $0.62\pm0.06$    &$0.011\pm0.003$       & 4/23\\
                 & 60--70\%     &$0.360\pm0.006$  & $0.350\pm0.006$  & $3700\pm160$ & $1920\pm110$   & $0.62\pm0.10$    &$0.005\pm0.0004$      & 19/23\\
                 & 70--80\%     &$0.350\pm0.006$  & $0.346\pm0.006$  & $3700\pm150$ & $1670\pm106$   & $0.62\pm0.08$    & $0.0024\pm0.0005$    & 11/23\\
\cline{2-8}
Fig. 1           & 0--5\%       &$0.542\pm0.009$  & $0.230\pm0.009$  & $4900\pm220$ & $2630\pm150$   & $0.985\pm0.13$   &$1.9\pm0.3$          & 11/20\\
Au-Au            & 5--10\%      &$0.529\pm0.007$  & $0.221\pm0.006$  & $4700\pm200$ & $2630\pm119$   & $0.993\pm0.05$   &$1.7\pm0.2$          & 10/23\\
27 GeV           & 10--20\%     &$0.522\pm0.008$  & $0.210\pm0.008$  & $4600\pm178$ & $2530\pm100$   & $0.98\pm0.08$    &$1.3\pm0.4$          & 8/20\\
                 & 20--30\%     &$0.510\pm0.006$  & $0.200\pm0.006$  & $4400\pm175$ & $2530\pm118$   & $0.99\pm0.09$    &$0.9\pm0.04$         & 4/20\\
                 & 30--40\%     &$0.490\pm0.008$  & $0.180\pm0.007$  & $4200\pm180$ & $2530\pm132$   & $0.98\pm0.11$    &$0.6\pm0.03$         & 8/20\\
                 & 40--50\%     &$0.478\pm0.009$  & $0.160\pm0.009$  & $4000\pm162$ & $2500\pm133$   & $0.99\pm0.08$    &$0.4\pm0.03$         & 14/20\\
                 & 50--60\%     &$0.460\pm0.007$  & $0.150\pm0.007$  & $4000\pm150$ & $2300\pm116$   & $0.99\pm0.10$    &$0.22\pm0.03$        & 20/20\\
                 & 60--70\%     &$0.404\pm0.006$  & $0.150\pm0.008$  & $4000\pm160$ & $2100\pm110$   & $0.99\pm0.12$    &$0.13\pm0.04$        & 11/20\\
                 & 70--80\%     &$0.364\pm0.008$  & $0.160\pm0.008$  & $3900\pm150$ & $2000\pm106$   & $0.99\pm0.09$    & $0.065\pm0.005$     & 19/20\\
\cline{2-8}
Fig. 1           & 0--5\%       &$0.542\pm0.005$  & $0.230\pm0.006$  & $5300\pm260$ & $2490\pm100$   & $0.985\pm0.06$   &$1.7\pm0.22$         & 14/19\\
Au-Au            & 5--10\%      &$0.529\pm0.005$  & $0.221\pm0.005$  & $5300\pm210$ & $2290\pm110$   & $0.993\pm0.07$   &$1.5\pm0.23$         & 11/19\\
39 GeV           & 10--20\%     &$0.522\pm0.007$  & $0.210\pm0.007$  & $5100\pm160$ & $2290\pm110$   & $0.98\pm0.08$    &$1.2\pm0.18$         & 10/19\\
                 & 20--30\%     &$0.510\pm0.006$  & $0.200\pm0.007$  & $5000\pm170$ & $2200\pm123$   & $0.99\pm0.11$    &$0.8\pm0.04$         & 10/19\\
                 & 30--40\%     &$0.490\pm0.008$  & $0.180\pm0.006$  & $5000\pm200$ & $2000\pm121$   & $0.98\pm0.04$    &$0.6\pm0.03$         & 7/19\\
                 & 40--50\%     &$0.480\pm0.005$  & $0.160\pm0.008$  & $4820\pm174$ & $2000\pm150$   & $0.99\pm0.09$    &$0.36\pm0.04$        & 1/19\\
                 & 50--60\%     &$0.460\pm0.007$  & $0.150\pm0.006$  & $4800\pm180$ & $1800\pm100$   & $0.99\pm0.06$    &$0.22\pm0.023$       & 5.5/19\\
                 & 60--70\%     &$0.404\pm0.008$  & $0.150\pm0.006$  & $4600\pm160$ & $1800\pm107$   & $0.99\pm0.13$    &$0.13\pm0.024$       & 19/19\\
                 & 70--80\%     &$0.364\pm0.008$  & $0.150\pm0.006$  & $4300\pm199$ & $1800\pm130$   & $0.99\pm0.08$    & $0.065\pm0.0022$    & 24/19\\
\cline{2-8}
Fig. 1           & 0--5\%       &$0.602\pm0.007$  & $0.430\pm0.006$  & $5700\pm240$ & $2900\pm130$   & $0.999\pm0.06$   &$13\pm1.3$           & 23/19\\
Au-Au            & 5--10\%      &$0.590\pm0.007$  & $0.430\pm0.006$  & $5200\pm200$ & $3200\pm130$   & $0.96\pm0.08$    &$6\pm0.4$            & 21/19\\
200 GeV          & 10--15\%     &$0.697\pm0.006$  & $0.413\pm0.006$  & $5300\pm190$ & $2900\pm130$   & $0.56\pm0.05$    &$2.5\pm0.3$          & 13/19\\
                 & 15--20\%     &$0.690\pm0.008$  & $0.413\pm0.007$  & $5300\pm180$ & $2700\pm135$   & $0.56\pm0.09$    &$1.4\pm0.3$          & 10/19\\
                 & 20--30\%     &$0.680\pm0.009$  & $0.410\pm0.006$  & $4900\pm200$ & $2800\pm128$   & $0.56\pm0.06$    &$0.55\pm0.04$        & 9/19\\
                 & 30--40\%     &$0.689\pm0.009$  & $0.413\pm0.007$  & $4800\pm174$ & $2730\pm150$   & $0.53\pm0.10$    &$0.25\pm0.04$        & 8/19\\
                 & 40--50\%     &$0.660\pm0.009$  & $0.404\pm0.008$  & $4600\pm180$ & $2730\pm140$   & $0.53\pm0.11$    &$0.14\pm0.02$        & 15/19\\
                 & 50--60\%     &$0.640\pm0.008$  & $0.402\pm0.007$  & $4400\pm160$ & $2700\pm137$   & $0.52\pm0.12$    &$0.09\pm0.004$       & 10/19\\
                 & 60--70\%     &$0.620\pm0.008$  & $0.404\pm0.008$  & $4560\pm199$ & $2400\pm138$   & $0.52\pm0.08$    & $0.043\pm0.005$     & 9/19\\
                 & 70--80\%     &$0.600\pm0.006$  & $0.389\pm0.009$  & $4300\pm160$ & $2400\pm126$   & $0.52\pm0.10$    &$0.019\pm0.004$      & 14/19\\
                 & 80--92\%     &$0.570\pm0.008$  & $0.370\pm0.007$  & $4000\pm199$ & $2500\pm100$   & $0.52\pm0.09$    & $0.01\pm0.005$      & 18/19\\
\hline
\end{tabular}%
\end{center}}
\end{table*}

\begin{table*}
{\scriptsize Table 2. Values of free parameters ($T_1$, $T_2$£¬
$V_1$, $V_2$ and $k$), normalization constant ($N_0$), $\chi^2$,
and degree of freedom (dof) corresponding to the curves in Fig. 2.
\vspace{-.50cm}
\begin{center}
\begin{tabular}{ccccccccccc}\\ \hline\hline
Collisions       & Centrality   & $T_1$ (GeV)     & $T_2$ (GeV) &
$V_1 (fm^3)$ & $V_2 (fm^3)$   & $k$              & $N_0$ &
$\chi^2$/ dof \\ \hline
Fig. 2           & 0--5\%       &$0.450\pm0.006$  & $0.895\pm0.008$  & $3000\pm200$ & $1730\pm130$   & $0.84\pm0.10$    &$4\pm0.5$           & 7/26\\
Au-Au            & 5--10\%      &$0.450\pm0.007$  & $0.795\pm0.008$  & $2900\pm160$ & $1730\pm125$   & $0.84\pm0.10$    &$3.7\pm0.3$         & 11/26\\
7.7 GeV          & 10--20\%     &$0.450\pm0.009$  & $0.765\pm0.008$  & $2643\pm190$ & $1706\pm108$   & $0.90\pm0.08$    &$2.75\pm0.3$        & 6/26\\
                 & 20--30\%     &$0.440\pm0.006$  & $0.705\pm0.007$  & $2543\pm160$ & $1643\pm128$   & $0.90\pm0.10$    &$1.9\pm0.2$         & 2.5/26\\
                 & 30--40\%     &$0.424\pm0.008$  & $0.605\pm0.007$  & $2400\pm170$ & $1578\pm120$   & $0.90\pm0.07$    &$1.4\pm0.2$         & 0.4/26\\
                 & 40--50\%     &$0.410\pm0.007$  & $0.470\pm0.007$  & $2300\pm170$ & $1551\pm110$   & $0.86\pm0.05$    &$0.8\pm0.05$        & 9/25\\
                 & 50--60\%     &$0.380\pm0.008$  & $0.280\pm0.006$  & $2240\pm160$ & $1430\pm100$   & $0.90\pm0.06$    &$0.5\pm0.04$        & 0.2/24\\
                 & 60--70\%     &$0.360\pm0.008$  & $0.320\pm0.006$  & $2000\pm169$ & $1300\pm100$   & $0.90\pm0.05$    &$0.3\pm0.04$        & 2.5/25\\
                 & 70--80\%     &$0.320\pm0.008$  & $0.150\pm0.005$  & $2000\pm170$ & $1200\pm80$    & $0.90\pm0.06$    &$0.14\pm0.03$       & 8/24\\
\cline{2-8}
Fig. 2           & 0--5\%       &$0.530\pm0.008$  & $0.200\pm0.008$  & $3300\pm201$ & $1780\pm110$   & $0.999\pm0.10$   &$2.8\pm0.4$         & 14/25\\
Au-Au            & 5--10\%      &$0.515\pm0.007$  & $0.200\pm0.007$  & $3200\pm197$ & $1780\pm100$   & $0.995\pm0.11$   &$2.2\pm0.4$         & 1.6/26\\
11.5 GeV         & 10--20\%     &$0.500\pm0.008$  & $0.190\pm0.007$  & $3100\pm170$ & $1720\pm100$   & $0.995\pm0.11$   &$1.8\pm0.4$         & 1/26\\
                 & 20--30\%     &$0.490\pm0.008$  & $0.210\pm0.008$  & $3000\pm150$ & $1683\pm80$    & $0.995\pm0.10$   &$1.2\pm0.3$         & 13/26\\
                 & 30--40\%     &$0.455\pm0.007$  & $0.224\pm0.007$  & $3000\pm134$ & $1500\pm80$    & $0.97\pm0.13$    &$0.8\pm0.04$        & 0.5/26\\
                 & 40--50\%     &$0.435\pm0.009$  & $0.224\pm0.007$  & $2960\pm161$ & $1400\pm96$    & $0.97\pm0.11$    &$0.5\pm0.04$        & 9/25\\
                 & 50--60\%     &$0.410\pm0.009$  & $0.224\pm0.005$  & $2800\pm152$ & $1400\pm93$    & $0.97\pm0.11$    &$0.3\pm0.03$        & 12/25\\
                 & 60--70\%     &$0.380\pm0.007$  & $0.200\pm0.006$  & $2760\pm150$ & $1300\pm85$    & $0.97\pm0.1$     &$0.15\pm0.03$       & 16/25\\
                 & 70--80\%     &$0.360\pm0.006$  & $0.180\pm0.007$  & $2500\pm120$ & $1300\pm99$    & $0.97\pm0.05$    &$0.08\pm0.003$      & 19/26\\
\cline{2-8}
Fig. 2           & 0--5\%       &$0.540\pm0.008$  & $0.230\pm0.007$  & $3500\pm170$ & $1800\pm120$   & $0.99\pm0.10$    &$2.3\pm0.3$         & 14/22\\
Au-Au            & 5--10\%      &$0.530\pm0.007$  & $0.217\pm0.006$  & $3300\pm190$ & $1834\pm132$   & $0.99\pm0.15$    &$2.1\pm0.2$         & 1/22\\
14.5 GeV         & 10--20\%     &$0.520\pm0.008$  & $0.210\pm0.006$  & $3200\pm150$ & $1758\pm140$   & $0.99\pm0.12$    &$1.5\pm0.2$         & 1.5/22\\
                 & 20--30\%     &$0.510\pm0.008$  & $0.200\pm0.007$  & $3200\pm119$ & $1658\pm135$   & $0.99\pm0.09$    &$1\pm0.06$          & 1.2/22\\
                 & 30--40\%     &$0.490\pm0.007$  & $0.180\pm0.007$  & $3000\pm140$ & $1650\pm119$   & $0.98\pm0.11$    &$0.7\pm0.02$        & 1/22\\
                 & 40--50\%     &$0.480\pm0.007$  & $0.160\pm0.007$  & $2800\pm152$ & $1668\pm108$   & $0.985\pm0.08$   &$0.4\pm0.03$        & 0.3/22\\
                 & 50--60\%     &$0.460\pm0.008$  & $0.150\pm0.008$  & $2700\pm120$ & $1568\pm100$   & $0.99\pm0.08$    &$0.2\pm0.03$        & 8/22\\
                 & 60--70\%     &$0.404\pm0.009$  & $0.150\pm0.007$  & $2600\pm106$ & $1500\pm99$    & $0.99\pm0.10$    &$0.12\pm0.02$       & 0.3/22\\
                 & 70--80\%     &$0.304\pm0.007$  & $0.150\pm0.009$  & $2400\pm80$  & $1500\pm76$    & $0.99\pm0.12$    & $0.065\pm0.004$    & 1/22\\
\cline{2-8}
Fig. 2           & 0--5\%       &$0.540\pm0.008$  & $0.230\pm0.007$  & $3700\pm170$ & $2000\pm120$   & $0.99\pm0.10$    &$2.1\pm0.3$         & 13/26\\
Au-Au            & 5--10\%      &$0.530\pm0.007$  & $0.217\pm0.006$  & $3700\pm190$ & $1834\pm132$   & $0.99\pm0.15$    &$1.9\pm0.22$        & 3/25\\
19.6 GeV         & 10--20\%     &$0.520\pm0.008$  & $0.210\pm0.006$  & $3600\pm150$ & $1758\pm140$   & $0.99\pm0.12$    &$1.45\pm0.21$       & 2/20\\
                 & 20--30\%     &$0.510\pm0.008$  & $0.200\pm0.007$  & $3500\pm119$ & $1658\pm135$   & $0.99\pm0.09$    &$0.92\pm0.04$       & 19/20\\
                 & 30--40\%     &$0.490\pm0.007$  & $0.180\pm0.007$  & $3300\pm140$ & $1650\pm119$   & $0.98\pm0.11$    &$0.65\pm0.024$      & 7/20\\
                 & 40--50\%     &$0.480\pm0.007$  & $0.160\pm0.007$  & $3100\pm152$ & $1668\pm108$   & $0.985\pm0.08$   &$0.4\pm0.03$        & 13/20\\
                 & 50--60\%     &$0.460\pm0.008$  & $0.150\pm0.008$  & $3100\pm120$ & $1568\pm100$   & $0.99\pm0.08$    &$0.23\pm0.025$      & 8/20\\
                 & 60--70\%     &$0.404\pm0.009$  & $0.150\pm0.007$  & $3000\pm106$ & $1500\pm99$    & $0.99\pm0.10$    &$0.13\pm0.015$      & 12/20\\
                 & 70--80\%     &$0.304\pm0.007$  & $0.150\pm0.009$  & $2900\pm80$  & $1400\pm76$    & $0.99\pm0.12$    & $0.68\pm0.04$      & 7/20\\
\cline{2-8}
Fig. 2           & 0--5\%       &$0.540\pm0.008$  & $0.230\pm0.007$  & $3500\pm170$ & $2800\pm120$   & $0.99\pm0.10$    &$2.1\pm0.3$         & 13/26\\
Au-Au            & 5--10\%      &$0.530\pm0.007$  & $0.217\pm0.006$  & $3300\pm190$ & $2800\pm132$   & $0.99\pm0.15$    &$1.9\pm0.22$        & 3/25\\
27 GeV           & 10--20\%     &$0.520\pm0.008$  & $0.210\pm0.006$  & $3205\pm150$ & $2758\pm140$   & $0.99\pm0.12$    &$1.45\pm0.21$       & 2/20\\
                 & 20--30\%     &$0.510\pm0.008$  & $0.200\pm0.007$  & $3200\pm119$ & $2600\pm135$   & $0.99\pm0.09$    &$0.92\pm0.04$       & 19/20\\
                 & 30--40\%     &$0.490\pm0.007$  & $0.180\pm0.007$  & $3000\pm140$ & $2500\pm119$   & $0.98\pm0.11$    &$0.65\pm0.024$      & 7/20\\
                 & 40--50\%     &$0.480\pm0.007$  & $0.160\pm0.007$  & $3000\pm152$ & $2300\pm108$   & $0.985\pm0.08$   &$0.4\pm0.03$        & 13/20\\
                 & 50--60\%     &$0.460\pm0.008$  & $0.150\pm0.008$  & $3000\pm120$ & $2200\pm100$   & $0.99\pm0.08$    &$0.23\pm0.025$      & 8/20\\
                 & 60--70\%     &$0.404\pm0.009$  & $0.150\pm0.007$  & $2500\pm106$ & $2500\pm99$    & $0.99\pm0.10$    &$0.13\pm0.015$      & 12/20\\
                 & 70--80\%     &$0.304\pm0.007$  & $0.150\pm0.009$  & $2400\pm80$  & $1500\pm76$    & $0.99\pm0.12$    & $0.68\pm0.04$      & 7/20\\
\cline{2-8}
Fig. 2           & 0--5\%       &$0.429\pm0.007$  & $0.469\pm0.006$  & $4000\pm260$ & $2900\pm100$   & $0.61\pm0.11$     &$1.5\pm0.2$          & 2/23\\
Au-Au            & 5--10\%      &$0.434\pm0.005$  & $0.420\pm0.005$  & $3934\pm210$ & $2900\pm110$   & $0.62\pm0.10$     &$0.65\pm0.05$        & 8/23\\
39 GeV           & 10--20\%     &$0.429\pm0.006$  & $0.420\pm0.008$  & $4000\pm160$ & $2672\pm110$   & $0.61\pm0.09$    &$0.25\pm0.04$         & 4/23\\
                 & 20--30\%     &$0.420\pm0.006$  & $0.420\pm0.008$  & $3800\pm170$ & $2700\pm123$   & $0.60\pm0.08$    &$0.12\pm0.03$         & 10/23\\
                 & 30--40\%     &$0.410\pm0.006$  & $0.410\pm0.006$  & $3900\pm200$ & $2450\pm121$   & $0.62\pm0.07$    &$0.07\pm0.004$        & 14/23\\
                 & 40--50\%     &$0.370\pm0.009$  & $0.400\pm0.006$  & $4000\pm174$ & $2000\pm150$   & $0.62\pm0.08$    &$0.03\pm0.003$        & 17/23\\
                 & 50--60\%     &$0.370\pm0.008$  & $0.370\pm0.006$  & $3800\pm180$ & $2000\pm100$   & $0.62\pm0.06$    &$0.011\pm0.003$       & 11/23\\
                 & 60--70\%     &$0.360\pm0.006$  & $0.350\pm0.006$  & $3600\pm160$ & $2000\pm107$   & $0.62\pm0.10$    &$0.005\pm0.0004$      & 29/23\\
                 & 70--80\%     &$0.350\pm0.006$  & $0.346\pm0.006$  & $3600\pm199$ & $1800\pm130$   & $0.62\pm0.08$    & $0.0024\pm0.0005$    & 26/23\\
\cline{2-8}
Fig. 2           & 0--5\%       &$0.489\pm0.009$  & $0.469\pm0.008$  & $5300\pm240$ & $2900\pm130$   & $0.63\pm0.13$    &$43\pm6$              & 6/13\\
Au-Au            & 5--10\%      &$0.479\pm0.008$  & $0.469\pm0.008$  & $4800\pm200$ & $3200\pm130$   & $0.63\pm0.10$    &$18\pm2$              & 15/13\\
200 GeV          & 10--15\%     &$0.469\pm0.008$  & $0.463\pm0.009$  & $4900\pm190$ & $3000\pm130$   & $0.635\pm0.09$   &$7\pm1$               & 13/13\\
                 & 15--20\%     &$0.459\pm0.008$  & $0.460\pm0.007$  & $4900\pm180$ & $2800\pm135$   & $0.635\pm0.09$   &$3.3\pm0.3$           & 2/13\\
                 & 20--30\%     &$0.450\pm0.006$  & $0.460\pm0.008$  & $4900\pm200$ & $2620\pm128$   & $0.62\pm0.07$    &$0.07\pm0.004$        & 14/23\\
                 & 30--40\%     &$0.450\pm0.009$  & $0.440\pm0.007$  & $4400\pm174$ & $2900\pm150$   & $0.62\pm0.09$    &$0.03\pm0.003$        & 17/23\\
                 & 40--50\%     &$0.448\pm0.009$  & $0.420\pm0.007$  & $4200\pm180$ & $2900\pm140$   & $0.62\pm0.06$    &$0.011\pm0.003$       & 11/23\\
                 & 50--60\%     &$0.426\pm0.007$  & $0.420\pm0.007$  & $4400\pm160$ & $2280\pm137$   & $0.62\pm0.12$    &$0.005\pm0.0004$      & 29/23\\
                 & 60--70\%     &$0.426\pm0.008$  & $0.430\pm0.008$  & $4400\pm199$ & $2230\pm138$   & $0.62\pm0.08$    & $0.0024\pm0.0005$    & 26/23\\
                 & 70--80\%     &$0.360\pm0.006$  & $0.420\pm0.009$  & $4300\pm160$ & $2100\pm126$   & $0.62\pm0.13$    &$0.005\pm0.0004$      & 29/23\\
                 & 80--92\%     &$0.420\pm0.008$  & $0.412\pm0.007$  & $4000\pm199$ & $2230\pm100$   & $0.62\pm0.09$    & $0.0024\pm0.0005$    & 26/23\\
\hline
\end{tabular}%
\end{center}}
\end{table*}

\begin{table*}
{\scriptsize Table 3. Values of free parameters ($T_1$, $T_2$£¬
$V_1$, $V_2$ and $k$), normalization constant ($N_0$), $\chi^2$,
and degree of freedom (dof) corresponding to the curves in Fig. 3.
\vspace{-.50cm}
\begin{center}
\begin{tabular}{ccccccccccc}\\ \hline\hline
Collisions       & Centrality   & $T_1$ (GeV)     & $T_2$ (GeV)      & $V_1 (fm^3)$ & $V_2 (fm^3)$   & $k$              & $N_0$              & $\chi^2$/ dof \\ \hline
Fig. 3           & 0--10\%      &$0.579\pm0.007$  & $0.295\pm0.008$  & $2000\pm150$ & $1500\pm80$    & $0.99\pm0.14$    &$0.027\pm0.004$     & 10/6\\
Au-Au            & 10--20\%     &$0.560\pm0.008$  & $0.105\pm0.006$  & $2400\pm110$ & $900\pm100$    & $0.99\pm0.10$    &$0.008\pm0.0006$    & 16/6\\
7.7 GeV          & 20--40\%     &$0.530\pm0.009$  & $0.535\pm0.008$  & $1800\pm90$  & $1280\pm100$   & $0.65\pm0.22$    &$0.0028\pm0.0005$   & 4/6\\
                 & 40--60\%     &$0.480\pm0.006$  & $0.240\pm0.009$  & $1700\pm100$ & $1200\pm80$    & $0.99\pm0.11$    &$4\times10^{-4}\pm6\times10^{-5}$ & 1.5/5\\
                 & 60--80\%     &$0.420\pm0.009$  & $0.355\pm0.008$  & $1600\pm160$ & $1270\pm40$    & $0.96\pm0.20$    &$4\times10^{-5}\pm5\times10^{-6}$  & 3/3\\
\cline{2-8}
 Au-Au           & 0--10\%     &$0.622\pm0.007$  & $0.205\pm0.007$  & $3200\pm170$ & $1100\pm70$     & $0.95\pm0.015$    &$0.007\pm0.0004$     & 7/6\\
 11.5 GeV        & 10--20\%    &$0.582\pm0.008$  & $0.195\pm0.007$  & $3100\pm125$ & $1000\pm75$     & $0.98\pm0.12$    &$0.0024\pm0.0005$  & 6/6\\
                 & 20--40\%   &$0.560\pm0.009$  & $0.200\pm0.006$  & $3000\pm110$ & $1000\pm100$      & $0.98\pm0.16$    &$7\times10^{-4}\pm6\times10^{-5}$    & 3/6\\
                 & 40--60\%   &$0.517\pm0.008$  & $0.205\pm0.009$  & $2000\pm100$ & $900\pm80$       & $0.97\pm0.13$    &$8.5\times10^{-5}\pm3\times10^{-6}$   & 2/5\\
                 & 60--80\%   &$0.440\pm0.009$ & $0.205\pm0.006$   & $2900\pm120$ & $700\pm50$       & $0.99\pm0.08$    &$7\times10^{-6}\pm4\times10^{-7}$    & 6/4\\
\cline{2-8}
  Au-Au          & 0--10\%    &$0.639\pm0.007$  & $0.755\pm0.008$  & $3300\pm160$ & $1600\pm90$      & $0.99\pm0.09$     &$0.004\pm0.0005$ & 9/6\\
  14.5 GeV      & 10--20\%   &$0.620\pm0.009$  & $0.612\pm0.009$  & $3200\pm110$ & $1500\pm90$      & $0.26\pm0.07$     &$8\times10^{-4}\pm4\times10^{-5}$ & 6/6\\
                 & 20--40\%   &$0.588\pm0.006$  & $0.605\pm0.008$  & $3400\pm120$ & $1120\pm100$    & $0.98\pm0.21$     &$3.5\times10^{-4}\pm6\times10^{-5}$  & 10/6\\
                 & 40--60\%   &$0.534\pm0.007$  & $0.631\pm0.009$  & $3000\pm90$ & $1300\pm90$      & $0.96\pm0.19$     &$6\times10^{-5}\pm5\times10^{-6}$  & 2/5\\
                 & 60--80\%   &$0.487\pm0.009$  & $0.455\pm0.008$  & $3000\pm90$ & $1000\pm70$      & $0.5\pm0.10$      &$3.3\times10^{-6}\pm5\times10^{-7}$ & 22/4\\
\cline{2-8}
  Au-Au          & 0--10\%    &$0.640\pm0.006$  & $0.600\pm0.008$  & $3300\pm130$ & $2200\pm90$      & $0.99\pm0.21$    &$0.0027\pm0.0004$   & 12/6\\
  19.6 GeV       & 10--20\%   &$0.614\pm0.008$  & $0.615\pm0.009$  & $3500\pm100$ & $1600\pm70$      & $0.96\pm0.14$    &$9\times10^{-4}\pm4\times10^{-5}$   & 11/6\\
                 & 20--40\%   &$0.588\pm0.007$  & $0.605\pm0.008$  & $3500\pm110$ & $1600\pm80$      & $0.97\pm0.17$    &$2.0\times10^{-4}\pm4\times10^{-5} $ & 5/6\\
                 & 40--60\%   &$0.535\pm0.006$  & $0.605\pm0.010$  & $3000\pm100$ & $1900\pm80$      & $0.95\pm0.10$    &$3.6\times10^{-5}\pm5\times10^{-6}$  & 2/5\\
                 & 60--80\%   &$0.468\pm0.007$  & $0.405\pm0.009$  & $2700\pm130$ & $1830\pm70$      & $0.99\pm0.15$    &$3.8\times10^{-6}\pm4\times10^{-7}$  & 13/4\\
\cline{2-8}
  Au-Au         & 0--10\%    &$0.641\pm0.007$  & $0.634\pm0.010$  & $3700\pm100$ & $2308\pm160$      & $0.88\pm0.16$    &$0.00135\pm0.0004$   & 6/6\\
  27 GeV        & 10--20\%   &$0.614\pm0.005$  & $0.650\pm0.010$  & $3600\pm110$ & $2250\pm180$      & $0.99\pm0.16$    &$6\times10^{-4}\pm5\times10^{-5}$  & 3/6\\
                 & 20--40\%   &$0.582\pm0.006$  & $0.605\pm0.009$ & $3000\pm115$ & $2715\pm180$      & $0.99\pm0.18$    &$2\times10^{-4}\pm4\times10^{-5}$  & 4/6\\
                 & 40--60\%   &$0.436\pm0.009$ & $0.605\pm0.007$  & $2900\pm130$ & $2500\pm170$      & $0.97\pm0.12$    &$3\times10^{-5}\pm4\times10^{-6}$  & 0.1/5\\
                 & 60--80\%   &$0.594\pm0.008$  & $0.453\pm0.008$ & $3000\pm120$ & $2236\pm190$      & $0.09\pm0.11$    &$7\times10^{-7}\pm3\times10^{-8}$  & 7/5\\
\cline{2-8}
  Au-Au         & 0--10\%   &$0.654\pm0.007$  & $0.305\pm0.009$  & $3500\pm106$ & $2908\pm194$       & $0.96\pm0.16$    &$0.0011\pm0.0004$  & 9/6\\
  39 GeV        & 10--20\%  &$0.618\pm0.008$  & $0.305\pm0.007$  & $3600\pm103$ & $2640\pm152$       & $0.99\pm0.13$    &$4.2\times10^{-4}\pm4\times10^{-5}$   & 4/6\\
                 & 20--40\%  &$0.588\pm0.009$  & $0.245\pm0.007$ & $3300\pm105$ & $2800\pm175$       & $0.98\pm0.14$    &$1.2\times10^{-4}\pm3\times10^{-5}$   & 6/6\\
                 & 40--60\%  &$0.540\pm0.010$  & $0.371\pm0.007$ & $3600\pm105$ & $2300\pm160$       & $0.99\pm0.15$    &$2\times10^{-5}\pm6\times10^{-6}$   & 0.6/6\\
                 & 60--80\%  &$0.465\pm0.008$  & $0.527\pm0.007$ & $3500\pm100$ & $2200\pm175$       & $0.99\pm0.13$    &$1.6\times10^{-6}\pm4\times10^{-7}$   & 13/5\\
\cline{2-8}
  Au-Au         & 0--10\%  &$0.690\pm0.007$  & $0.505\pm0.008$  & $3800\pm120$ & $3200\pm190$       & $0.99\pm0.14$    &$8.5\times10^{-4}\pm4\times10^{-5}$ & 5/6\\
  62.4 GeV       & 10--20\% &$0.663\pm0.008$  & $0.604\pm0.010$ & $4000\pm240$ & $2830\pm180$       & $0.96\pm0.13$    &$2.6\times10^{-4}\pm5\times10^{-5}$ & 4/6\\
                 & 20--40\% &$0.630\pm0.009$  & $0.463\pm0.008$ & $3700\pm100$ & $2900\pm90$        & $0.98\pm0.10$    &$8\times10^{-5}\pm4\times10^{-6}$   & 6/6\\
                 & 40--60\% &$0.590\pm0.007$  & $0.532\pm0.009$ & $3400\pm115$ & $3000\pm85$        & $0.97\pm0.20$    &$1.4\times10^{-5}\pm4\times10^{-6}$ & 22/6\\
                 & 60--80\% &$0.515\pm0.007$  & $0.297\pm0.010$ & $4000\pm110$ & $2130\pm70$        & $0.96\pm0.17$    &$1.1\times10^{-6}\pm5\times10^{-7}$ & 6/5\\
\cline{2-8}
  Au-Au         & 0--10\%   &$0.800\pm0.008$  & $0.434\pm0.009$ & $4300\pm100$ & $3300\pm200$       & $0.96\pm0.21$    &$5.8\times10^{-4}\pm4\times10^{-5}$   & 12/6\\
  200 GeV       & 10--20\%  &$0.745\pm0.007$  & $0.932\pm0.008$ & $4000\pm110$ & $3400\pm210$       & $0.97\pm0.12$    &$2.3\times10^{-4}\pm5\times10^{-5}$   & 5/6\\
                & 20--40\%  &$0.700\pm0.008$ & $0.576\pm0.009$  & $4800\pm120$ & $2430\pm180$       & $0.95\pm0.17$    &$6.8\times10^{-5}\pm5\times10^{-6} $  & 15/6\\
                & 40--40\%  &$0.650\pm0.009$  & $0.679\pm0.008$ & $4100\pm100$ & $2900\pm190$       & $0.93\pm0.15$    &$1.25\times10^{-5}\pm4\times10^{-6}$  & 1/6\\
                & 60--80\%  &$0.550\pm0.010$  & $0.705\pm0.007$ & $4000\pm100$ & $2800\pm80$        & $0.96\pm0.13$    &$1.4\times10^{-6}\pm3\times10^{-7}$   & 2/5\\
\hline
\end{tabular}%
\end{center}}
\end{table*}

Figure 6 is similar to figure 5, but it shows the dependence of
kinetic freeze-out volume on energy and centrality. The kinetic
freeze-out volume increases with energy due to larger initial bulk
system at higher energies. The increase in energy leads to long
evolution time which corresponds to larger partonic system and the
kinetic freeze-out volume becomes larger in large partonic system.
We want to point out that here the initial system is not for
the system at evolution time being 0, but at the moment after
initial collision, which means that the evolution time is greater
than 0, says 1 $fm/c$ for example, for the initial system. The
higher the collision energy, the larger the initial system. At
present, there is no observation of critical volume but we can
study it in future, by analyzing more higher energies and
different collision systems containing the production of different
particles. In addition, the kinetic freeze-out volume also
decreases with decreasing centrality. In fact, there are large
number of binary collisions in central collisions due to
re-scattering of partons which leads the system quickly to
equilibrium state. While when centrality decreases, the number of
participants decreases, which leads the system to equilibrium
state in a steady manner. We also observed that the kinetic
freeze-out volume is mass dependent. The lighter the particle, the
larger the kinetic freeze-out volume. This observation shows the
early freeze-out of the heavier particles compared to the lighter
ones, and this may indicate different freeze-out surfaces for
various particles. One can found such results in literature [56,
57].

\begin{figure*}[htbp]
\begin{center}
\includegraphics[width=14.cm]{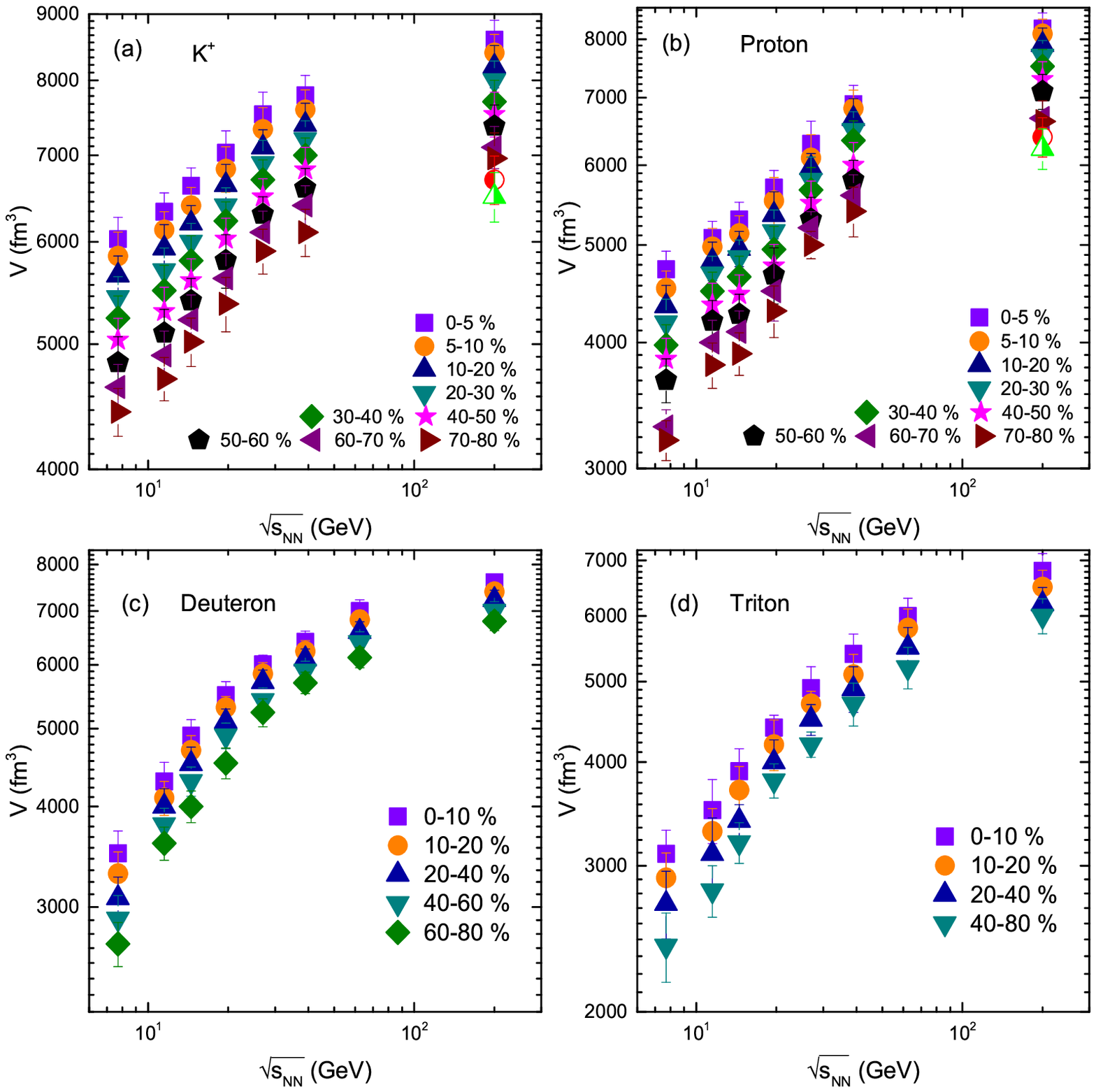}
\end{center}
Fig. 6. Similar to fig. 5, but it presents the dependence of
kinetic freeze-out volume on energy and centrality for $K^+$, $p$,
$d$ and $t$.
\end{figure*}
\begin{table*}
{\scriptsize Table 4. Values of free parameters ($T_1$, $T_2$£¬
$V_1$, $V_2$ and $k$), normalization constant ($N_0$), $\chi^2$,
and degree of freedom (dof) corresponding to the curves in Fig. 4.
\vspace{-.50cm}
\begin{center}
\begin{tabular}{ccccccccccc}\\ \hline\hline
Collisions       & Centrality   & $T_1$ (GeV)     & $T_2$ (GeV)
& $V_1 (fm^3)$ & $V_2 (fm^3)$   & $k$       & $N_0$              &
$\chi^2$/ dof \\ \hline
Fig. 4           & 0--10\%      &$0.691\pm0.005$  & $0.595\pm0.006$  & $2000\pm130$ & $1100\pm80$    & $0.99\pm0.24$    &$1.2\times10^{-4}\pm3\times10^{-5}$& 7/2\\
Au-Au            & 10--20\%     &$0.657\pm0.006$  & $0.595\pm0.005$  & $1800\pm108$ & $1100\pm100$   & $0.97\pm0.20$    &$3.9\times10^{-5}\pm4\times10^{-6}$& 4/2\\
7.7 GeV          & 20--40\%     &$0.640\pm0.007$  & $0.585\pm0.006$  & $1600\pm143$ & $1100\pm110$   & $0.94\pm0.22$    &$9.7\times10^{-6}\pm5\times10^{-7}$& 5/2\\
                 & 40--60\%     &$0.650\pm0.006$  & $0.450\pm0.008$  & $1300\pm120$ & $1100\pm110$   & $0.11\pm0.02$    &$2.4\times10^{-6}\pm4\times10^{-7}$& 2/0\\
\cline{2-8}
 Au-Au           & 0--10\%     &$0.200\pm0.006$  & $0.830\pm0.006$  & $3000\pm150$ & $1500\pm160$    & $0.19\pm0.01$    &$3.9\times10^{-5}\pm4\times10^{-6}$& 2/2\\
 11.5 GeV        & 10--20\%    &$0.200\pm0.005$  & $0.710\pm0.006$  & $2000\pm100$ & $1300\pm115$    & $0.03\pm0.006$   &$1.5\times10^{-5}\pm3\times10^{-6}$& 4/2\\
                 & 20--40\%   &$0.650\pm0.007$   & $0.664\pm0.007$  & $1800\pm120$ & $1300\pm105$    & $0.98\pm0.26$    &$1.9\times10^{-6}\pm6\times10^{-7}$ & 6/\\
                 & 40--80\%   &$0.540\pm0.005$  & $0.470\pm0.007$   & $1700\pm90 $ & $1100\pm110$    & $0.80\pm0.23$    &$4.2\times10^{-7}\pm6\times10^{-8}$ & 1/1\\
\cline{2-8}
  Au-Au          & 0--10\%    &$0.767\pm0.006$  & $0.295\pm0.007$  & $2200\pm140$ & $1700\pm110$    & $0.99\pm0.25$     &$0.0013\pm0.0004$                  & 6/2\\
  14.5 GeV      & 10--20\%    &$0.739\pm0.004$  & $0.295\pm0.009$  & $2400\pm150$ & $1300\pm100$    & $0.99\pm0.30$     &$3.6\times10^{-6}\pm5\times10^{-7}$& 4/2\\
                 & 20--40\%   &$0.700\pm0.006$  & $0.340\pm0.006$  & $2000\pm135$ & $1400\pm120$    & $0.99\pm0.27$     &$9.6\times10^{-7}\pm3\times10^{-8}$ & 32/2\\
                 & 40--80\%   &$0.460\pm0.006$  & $0.525\pm0.006$  & $2000\pm90$ & $1200\pm90$      & $0.38\pm0.19$     &$2.7\times10^{-7}\pm5\times10^{-8}$  & 10/2\\
\cline{2-8}
  Au-Au          & 0--10\%    &$0.767\pm0.006$  & $0.295\pm0.007$  & $3200\pm135$ & $1200\pm120$    & $0.99\pm0.25$     &$4.7\times10^{-6}\pm6\times10^{-7}$ & 3/2\\
  19.6 GeV       & 10--20\%   &$0.740\pm0.004$  & $0.295\pm0.009$  & $3000\pm170$ & $1200\pm126$    & $0.99\pm0.30$     &$1.5\times10^{-6}\pm5\times10^{-7}$ & 5/2\\
                 & 20--40\%   &$0.696\pm0.006$  & $0.340\pm0.006$  & $3000\pm140$ & $1000\pm114$    & $0.99\pm0.27$     &$3.3\times10^{-7}\pm7\times10^{-8}$ & 6/2\\
                 & 40--80\%   &$0.650\pm0.007$  & $0.425\pm0.009$  & $2000\pm90$  & $1800\pm90$     & $0.58\pm0.19$     &$9.3\times10^{-8}\pm4\times10^{-9}$ & 1/2\\
\cline{2-8}
  Au-Au         & 0--10\%    &$0.910\pm0.005$  & $0.205\pm0.006$  & $3000\pm150$ & $1910\pm150$    & $0.79\pm0.20$      &$2.7\times10^{-6}\pm4\times10^{-7}$ & 4/2\\
  27 GeV        & 10--20\%   &$0.740\pm0.006$  & $0.293\pm0.007$  & $2600\pm135$ & $2100\pm133$    & $0.99\pm0.18$      &$8.6\times10^{-7}\pm5\times10^{-8}$ & 4/2\\
                 & 20--40\%   &$0.696\pm0.006$  & $0.340\pm0.006$  & $3400\pm150$ & $1100\pm142$   & $0.99\pm0.22$      &$2.2\times10^{-7}\pm6\times10^{-8}$ & 3/2\\
                 & 40--80\%   &$0.579\pm0.005$  & $0.425\pm0.007$  & $3200\pm120$ & $1000\pm128$   & $0.85\pm0.19$      &$4\times10^{-8}\pm5\times10^{-9}$ & 0.6/1\\
\cline{2-8}
  Au-Au         & 0--10\%   &$0.928\pm0.007$  & $0.291\pm0.008$   & $3000\pm154$ & $2400\pm160$    & $0.74\pm0.22$      &$1.5\times10^{-6}\pm4\times10^{-7}$ & 12/2\\
  39 GeV        & 10--20\%   &$0.740\pm0.007$  & $0.293\pm0.008$  & $3000\pm150$ & $2100\pm142$    & $0.99\pm0.18$      &$4.7\times10^{-7}\pm5\times10^{-8}$ & 5/2\\
                 & 20--40\%   &$0.696\pm0.006$  & $0.325\pm0.006$ & $3200\pm160$ & $1700\pm150$    & $0.99\pm0.20$      &$1.2\times10^{-7}\pm4\times10^{-8}$ & 4/2\\
                 & 40--80\%   &$0.715\pm0.005$  & $0.410\pm0.007$ & $2500\pm150$ & $2200\pm130$    & $0.45\pm0.10$     &$3.3\times10^{-8}\pm7\times10^{-9}$  & 2/2\\
\cline{2-8}
  Au-Au         & 0--10\%  &$0.260\pm0.006$   & $0.880\pm0.006$ & $3200\pm140$ & $2800\pm160$       & $0.13\pm0.04$    &$8\times10^{-7}\pm6\times10^{-8}$   & 8/0\\
  62.4 GeV       & 10--20\% &$0.780\pm0.005$  & $0.267\pm0.008$ & $3000\pm160$ & $2800\pm157$      & $0.99\pm0.23$     &$2.6\times10^{-7}\pm6\times10^{-8}$ & 6/0\\
                 & 20--40\% &$0.740\pm0.007$  & $0.405\pm0.006$ & $3000\pm140$ & $2500\pm170$       & $0.98\pm0.20$    &$8.2\times10^{-8}\pm5\times10^{-9}$ & 13/0\\
                 & 40--80\% &$0.700\pm0.008$  & $0.400\pm0.006$ & $3200\pm150$ & $2000\pm150$       & $0.99\pm0.28$    &$1.6\times10^{-8}\pm4\times10^{-9}$ & 5/0\\
\cline{2-8}
  Au-Au         & 0--10\%   &$0.925\pm0.009$  & $0.289\pm0.009$ & $4000\pm160$ & $2800\pm170$       & $0.99\pm0.24$    &$2.4\times10^{-7}\pm6\times10^{-8}$    & 1/-\\
  200 GeV       & 10--20\%  &$0.892\pm0.008$  & $0.276\pm0.006$ & $4000\pm170$ & $2500\pm140$       & $0.99\pm0.17$    &$1.1\times10^{-7}\pm5\times10^{-8}$  & 3/-\\
                & 20--40\%  &$0.812\pm0.007$  & $0.250\pm0.007$ & $4000\pm150$ & $2200\pm140$       & $0.98\pm0.23$    &$3.6\times10^{-8}\pm5\times10^{-9}$   & 4/-\\
                & 40--80\%  &$0.700\pm0.007$  & $0.995\pm0.007$ & $4000\pm150$ & $2000\pm140$       & $0.90\pm0.23$    &$8\times10^{-9}\pm4\times10^{-10}$     & 9/-\\
\hline
\end{tabular}%
\end{center}}
\end{table*}

Figure 7 and 8 are also similar to figure 5 and 6, but they show
the energy and centrality dependent mean transverse momentum
($<p_T>$) and initial temperature ($T_i$) respectively. We have
analyzed the mean $p_T$ ($<p_T>$) and the root-mean-square $p_T$
($\sqrt{<p^2_T>}$) over $\sqrt2$ ($\sqrt{<p^2_T>}$/$\sqrt2$) which
are calculated from the fit function over a given $p_T$ range
and it is noteworthy that $p_T$ used to obtain the initial
temperature and mean $p_T$ is extrapolated up to zero. According
to string percolation model [58--60], ($\sqrt{<p^2_T>}$/$\sqrt2$)
represent the initial temperature ($T_i$). One can see that that
$<p_T>$ and $T_i$ increases with energy from 7.7 to 14.5 and then
remains unchanged up to 39 GeV and then again rises at 62.4 and
200 GeV. Like $T$ and $V$, both $<p_T>$ and $T_i$ increases from
peripheral to central collisions. Both $<p_T>$ and $T_i$ are
mass dependent. Heavier the particle, larger the $<p_T>$ and $T_i$.

Although the parameters $T$, $V$, $T_i$ and $T_0$ are extracted
from the model, in fact they are based on the fit to experimental
data. In ref. [61, 62] $T_i$ does not depend on the model, but is
equal to $\sqrt{<p^2_T>}$/$\sqrt2$ and can be obtained from the
$p_T$ spectra, and we can also obtain it from the data directly.

\begin{figure*}[htbp]
\begin{center}
\includegraphics[width=14.cm]{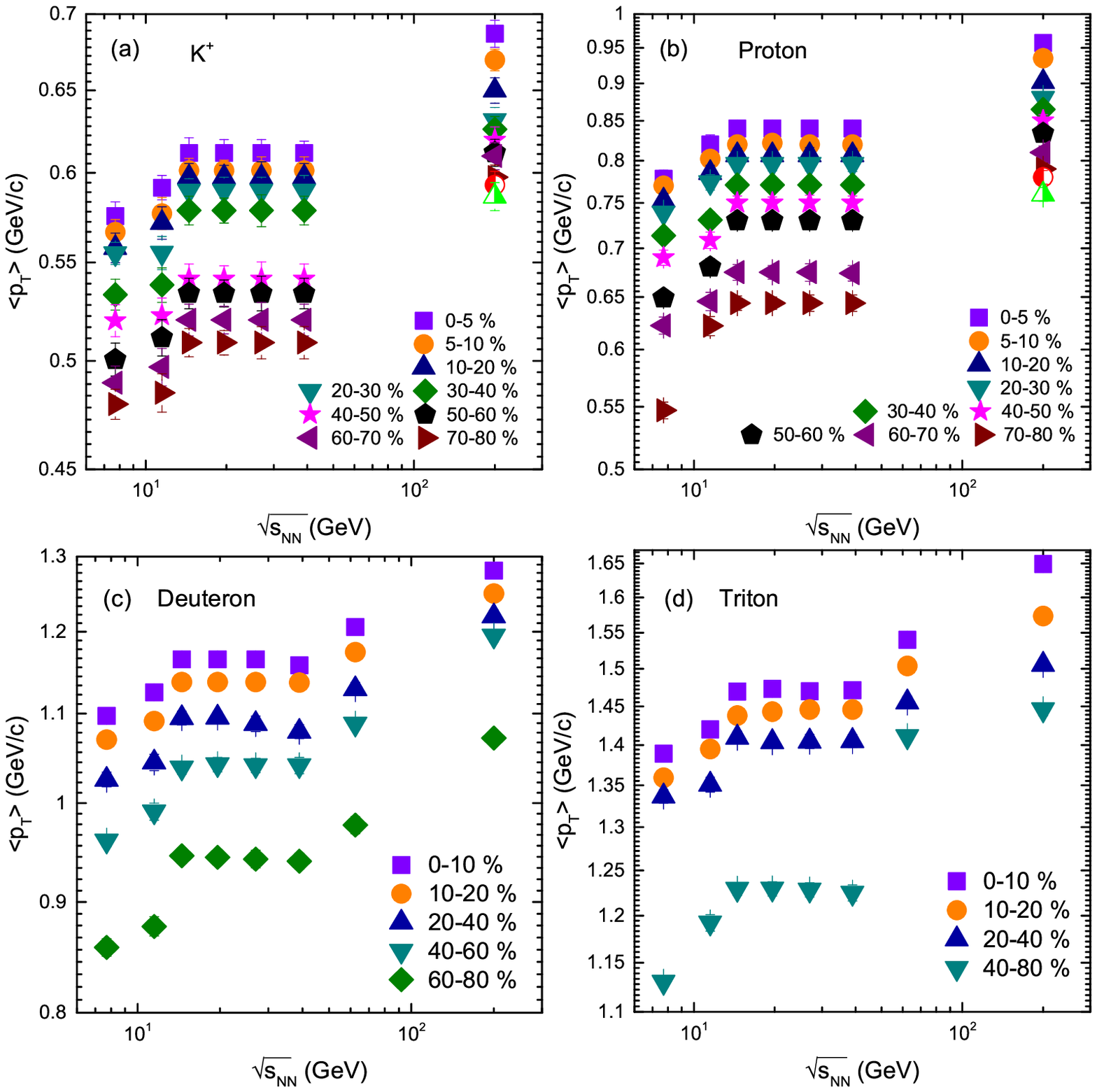}
\end{center}
Fig. 7. Similar to fig. 5, but it presents the dependence of mean
transverse momentum on energy and centrality for $K^+$, $p$, $d$
and $t$.
\end{figure*}

\begin{figure*}[htbp]
\begin{center}
\includegraphics[width=14.cm]{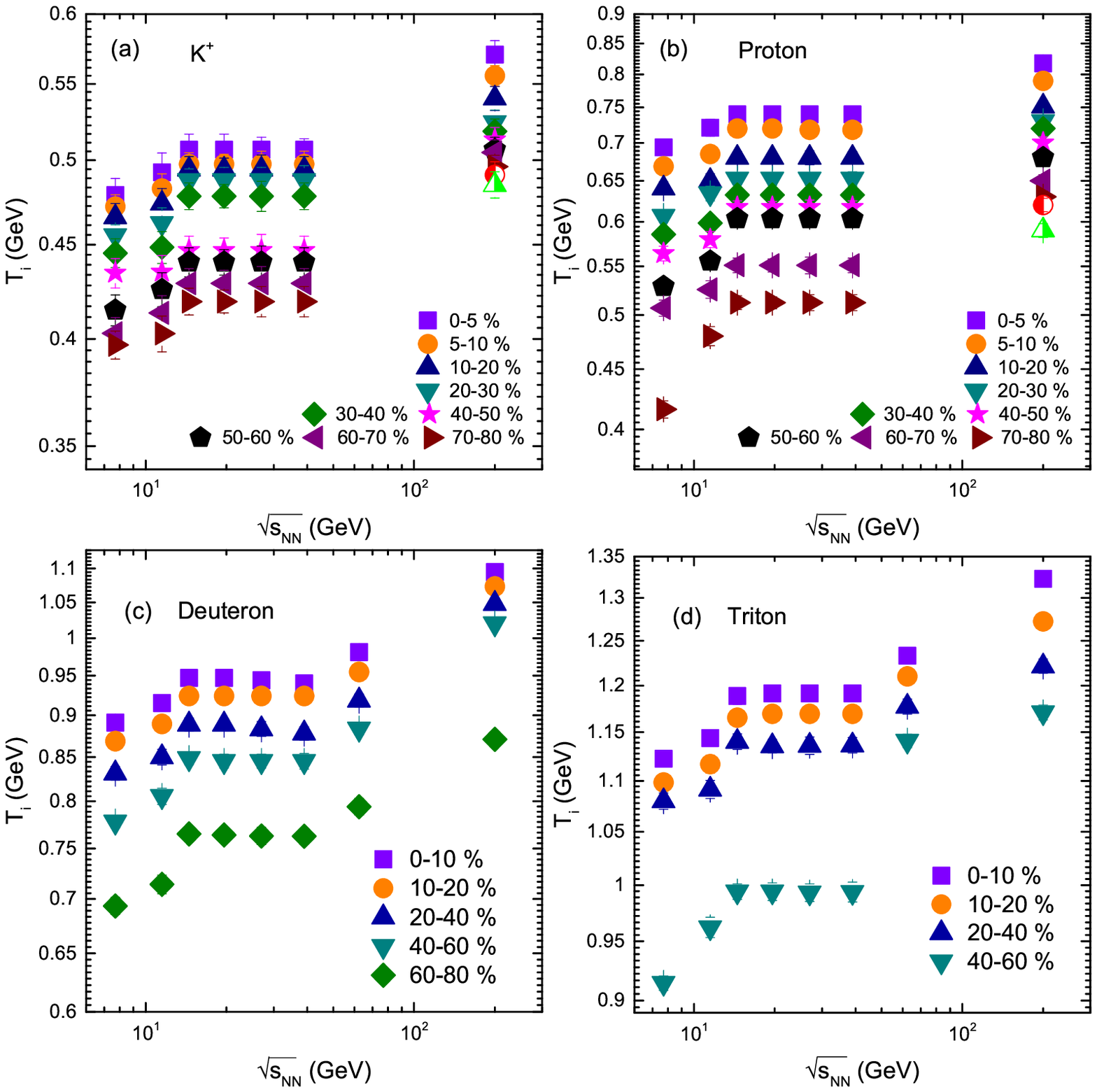}
\end{center}
Fig. 8. Similar to fig. 5, but it presents the dependence of
initial temperature on energy and centrality for $K^+$, $p$, $d$
and $t$.
\end{figure*}

\begin{figure*}[htbp]
\begin{center}
\includegraphics[width=14.cm]{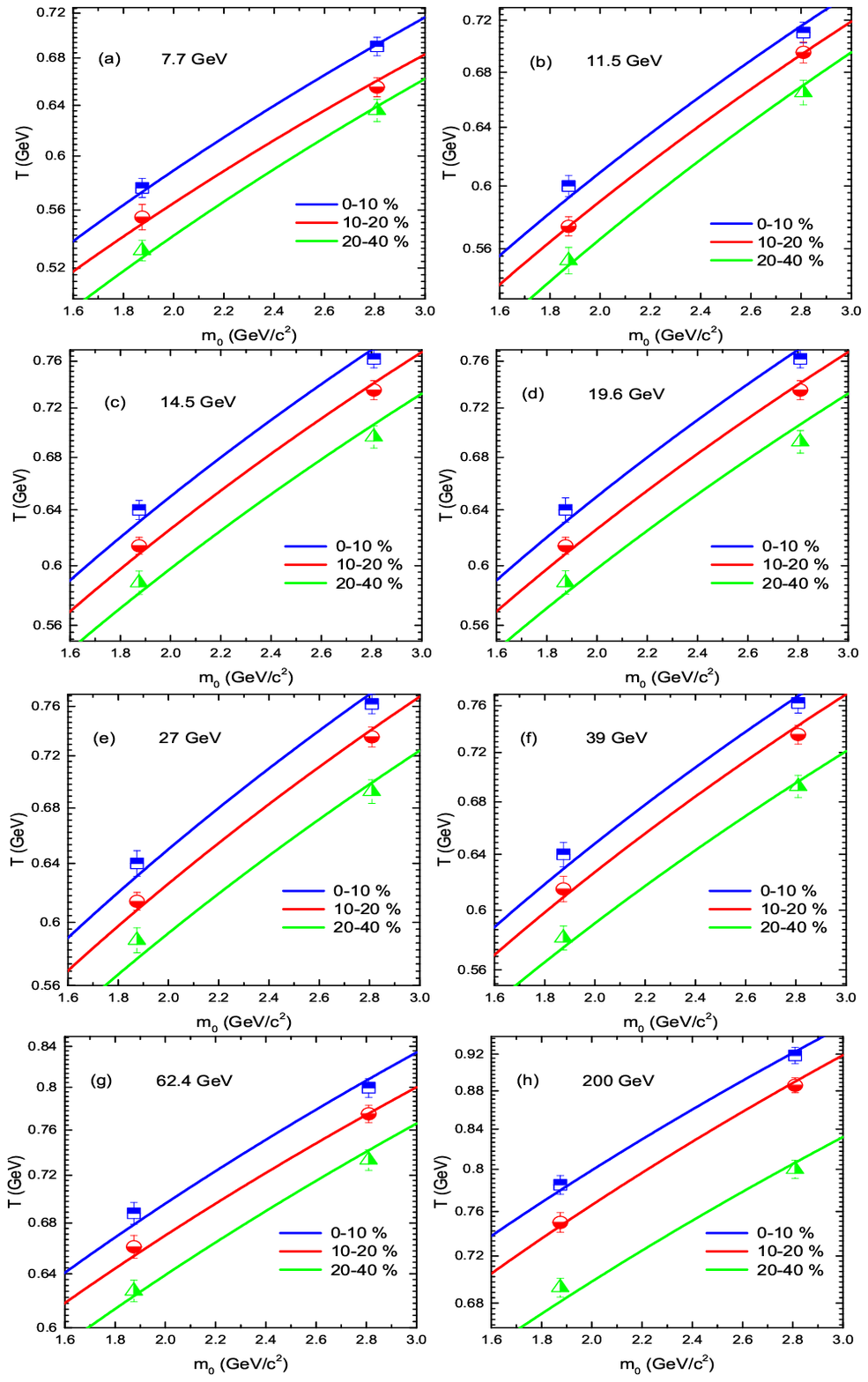}
\end{center}
Fig. 9. Relation between the effective temperature and $m_0$.
\end{figure*}
Figure 9 shows the relation between the effective temperature and
rest mass of the particle. Deuteron and triton at different
energies are put together in different panels and are fitted by
the linear equation by the relation $T=T_0+am_0$, where $T_0$ is
the intercept and $m_0$ is the slope. The intercepts of the
emission source is $T_0$ for massless particles e;g the intercepts
starts from $m_0$=0. Therefore, it does not matter whether two or
three data points. The intercepts from two, three and five data
points may be different, that indicates the multiple temperature,
or different stages or different times of formation of different
particles. The intercepts and slopes with the value of $\chi^2$
obtained by the linear function are listed in table 3. The
effective temperature versus rest mass of the particle at 7.7,
11.5, 14.5, 19.6, 27, 39, 62.4 and 200 GeV are presented in panels
(a), (b), (c), (d), (e), (f), (g) and (h) respectively in 0--10\%,
10--20\% and 20--40\% centrality bins. The last centrality classes
are missing because they are different for deuteron and triton
(40--60\% and 60--80\% for deuteron, and 40--80\% for triton). In
order to include them in the present work, we have to distribute
40--80\% centrality bin of triton, and it is not an easy job, so
we skip it and showed the first three centrality classes. It
should be noted that we also skipped to include $K^+$ and $p$ in
the above relation because of their different centrality intervals
from $d$ and $t$. In fact we only presented a few centrality
classes of $d$ and $t$ which were the same, in order to show that
how to extract $T_0$ by the alternative method. In fig. 7, the
intercepts between the effective temperatures and the rest mass of
the particles are regarded as the average kinetic freeze-out
temperature [9, 16].
\begin{table*}
\vskip.3cm {\small Table 5. Values of slopes and intercepts in
different types of collisions at the RHIC. \vspace{-.20cm}
\begin{center}
\begin{tabular}{cccccc}\\ \hline\hline
Collisions & $\sqrt{s_{NN}}$ ($\sqrt{s}$) & centrality & $slope$
&$intercept$ & $\chi^2$ \\ \hline
  Au-Au    & 7.7 GeV & 0--10\% &$0.127\pm0.007$ & $0.335\pm0.007$ & $0.2 $\\
           &          & 10--20\% &$0.118\pm0.008$ & $0.323\pm0.006$ & $0.7 $\\
           &          & 20--40\% &$0.120\pm0.008$ & $0.302\pm0.008$ & $0.4 $\\
\hline
Au-Au    & 11.5 GeV & 0--10\% &$0.133\pm0.008$ & $0.343\pm0.008$ & $1.7 $\\
          &           & 10--20\% &$0.129\pm0.006$ & $0.332\pm0.009$& $0.006 $\\
          &           & 20--40\% &$0.126\pm0.008$ & $0.308\pm0.006$& $0.5 $\\
\hline
Au-Au    & 14.5 GeV    & 0--10\% &$0.150\pm0.007$ & $0.350\pm0.007$ & $2 $\\
          &           & 10--20\% &$0.142\pm0.007$ & $0.342\pm0.007$ & $0.9 $\\
          &           & 20--40\% &$0.134\pm0.007$ & $0.330\pm0.008$ & $2$\\
\hline
Au-Au    & 19.6 GeV   & 0--10\% &$0.150\pm0.006$ & $0.350\pm0.008$ & $2 $\\
          &           & 10--20\% &$0.142\pm0.007$ & $0.342\pm0.005$ & $0.9$\\
          &           & 20--40\% &$0.134\pm0.007$ & $0.330\pm0.008$ & $3$\\
\hline
Au-Au    & 27 GeV     & 0--10\% &$0.150\pm0.007$ & $0.350\pm0.006$ & $2 $\\
          &           & 10--20\% &$0.142\pm0.006$ & $0.342\pm0.006$ & $0.9$\\
          &           & 20--40\% &$0.134\pm0.007$ & $0.331\pm0.007$ & $0.8$\\
\hline
Au-Au    & 39 GeV     & 0--10\% &$0.150\pm0.008$ & $0.350\pm0.007$ & $2 $\\
          &           & 10--20\% &$0.143\pm0.006$ & $0.341\pm0.006$ & $0.1$\\
          &           & 20--40\% &$0.130\pm0.007$ & $0.331\pm0.007$ & $0.8$\\
\hline
Au-Au    & 62.4 GeV   & 0--10\% &$0.138\pm0.009$ & $0.420\pm0.006$  & $2 $\\
          &           & 10--20\% &$0.130\pm0.008$ & $0.410\pm0.007$ & $0.8$\\
          &           & 20--40\% &$0.127\pm0.008$ & $0.385\pm0.009$ & $1$\\
\hline
Au-Au     & 200 GeV   & 0--10\% &$0.153\pm0.007$ & $0.493\pm0.006$  & $0.6$\\
          &           & 10--20\% &$0.153\pm0.008$ & $0.460\pm0.008$ & $0.3$\\
          &           & 20--40\% &$0.134\pm0.007$ & $0.430\pm0.008$ & $0.7$\\
\hline
\end{tabular}
\end{center}}
\end{table*}

Figure 10 shows the dependence of kinetic freeze-out temperature.
The symbols from up to downward and from left to right shows the
dependence of kinetic freeze-out temperature on centrality and
energy respectively. We noticed three trends of $T_0$ with rising
the collision energy, (1) from 7.7 to 14.5 GeV which is
increasing, (2) 14.5 to 39 GeV which remains consistent, and (3)
again increasing trend, which indicate towards different
interaction mechanisms in $\sqrt{s_{NN}}$=14.5-39 GeV,
$\sqrt{s_{NN}}$ $<$ 14.5 GeV and $\sqrt{s_{NN}}$ $>$ 39 GeV. We
believe that if, the baryon-dominated effect plays more important
role at below 14.5 GeV in AA collisions, the meson-dominated
effect should play more important role at above 14.5 GeV. In case
of baryon-domination, the system deposits less energy, and then
the system has low excitation degree and temperature, However the
case is opposite in meson dominated case. In fact, 14.5 GeV is a
particular energy that may need more attention to be paid. It
seems that the phase transition occurs in part volume from 14.5
GeV to 39 GeV, where 14.5 GeV may be the onset energy of the part
phase transition and 39 GeV is the whole phase transition. The
range of critical energy is 14.5 GeV to 39 GeV. If the region from
14.5 to 39 GeV in the excitation functions of $T_0$ is regarded as
a reflection of the formation of QGP liquid drop, the quick rise
of $T_0$ at 62.4 and 200 GeV is a reflection of higher temperature
QGP liquid drop due to larger energy deposition. At 62.4 and 200
GeV, the higher collision energy should result in creation of
larger energy density and blast wave, and then higher $T_0$.
It should be noted that, the above discussion on the excitation
function of $T_0$ presented in Fig. 10 is also suitable to the
excitation function of $T$ presented in Fig. 5 and $T_i$ in fig.
8, though the effect of flow is included in the former. We believe
that $T_i$, $T$ and $T_0$ has the same trend because the later two
are the reflection of the former. We would like to specify that we
included $K^+$ and $p$ with $d$ and $t$ in order to check the
difference in trend of the parameters of $K^+$ and $p$ from $d$
and $t$. There is no difference observed in the trend of
parameters for these particles. Different particle species can
give information about the differences in their emission but they
are not responsible to cause any difference in the trend of the
parameters. In deed, different models and methods used in the
extraction of the parameters can give their different trend. Such
studies are present in literature [37, 52, 63, 64, 65].
Furthermore, it can also be seen that the temperature of $t$ is
the largest and that for $K^+$ is the lowest which shows that
heavier particles decouple early from the system. The effective
temperature, initial temperature and kinetic freeze-out
temperature has the same behavior with increasing energy, and
hence their explanation is the same. Furthermore, like the
effective and initial temperature, the kinetic freeze-out
temperature also increases with centrality which indicates
that the central collision is more harsh and the energy deposited
in central collisions system is larger compared to the peripheral
collisions, and hence it results in larger temperature. The larger
$T$ ($T$ or $T_0$) in central collisions means that the particles
in central collisions decouple from the system more early compared
to the peripheral collisions.

In addition, we observed that the initial temperature is larger
than the effective temperature, and the later is larger than the
kinetic freeze-out temperature. In general, the chemical
freeze-out temperature occurs between the initial and kinetic
freeze-out temperatures, and is roughly equal to the effective
temperature. This sequence is consistent with the order of time
evolution of the interacting system, although both the effective
and kinetic freeze-out temperatures are extracted at the
kinetic/thermal freeze-out. However, the four temperatures can not
be compared directly in most cases due to different thermometric
scales being used. For example, one use thermometric scales in
thermal and statistical physics, a method is needed to unify
different thermometric scales in subatomic physics, but it is
beyond the focus of the present work to structure the method. This
issue shall not be discussed further in the present work. It is
noteworthy that the values of temperatures in the present work are
higher than the normal values because different scale is used in
their extraction.
\begin{figure*}[htbp]
\begin{center}
\includegraphics[width=14.cm]{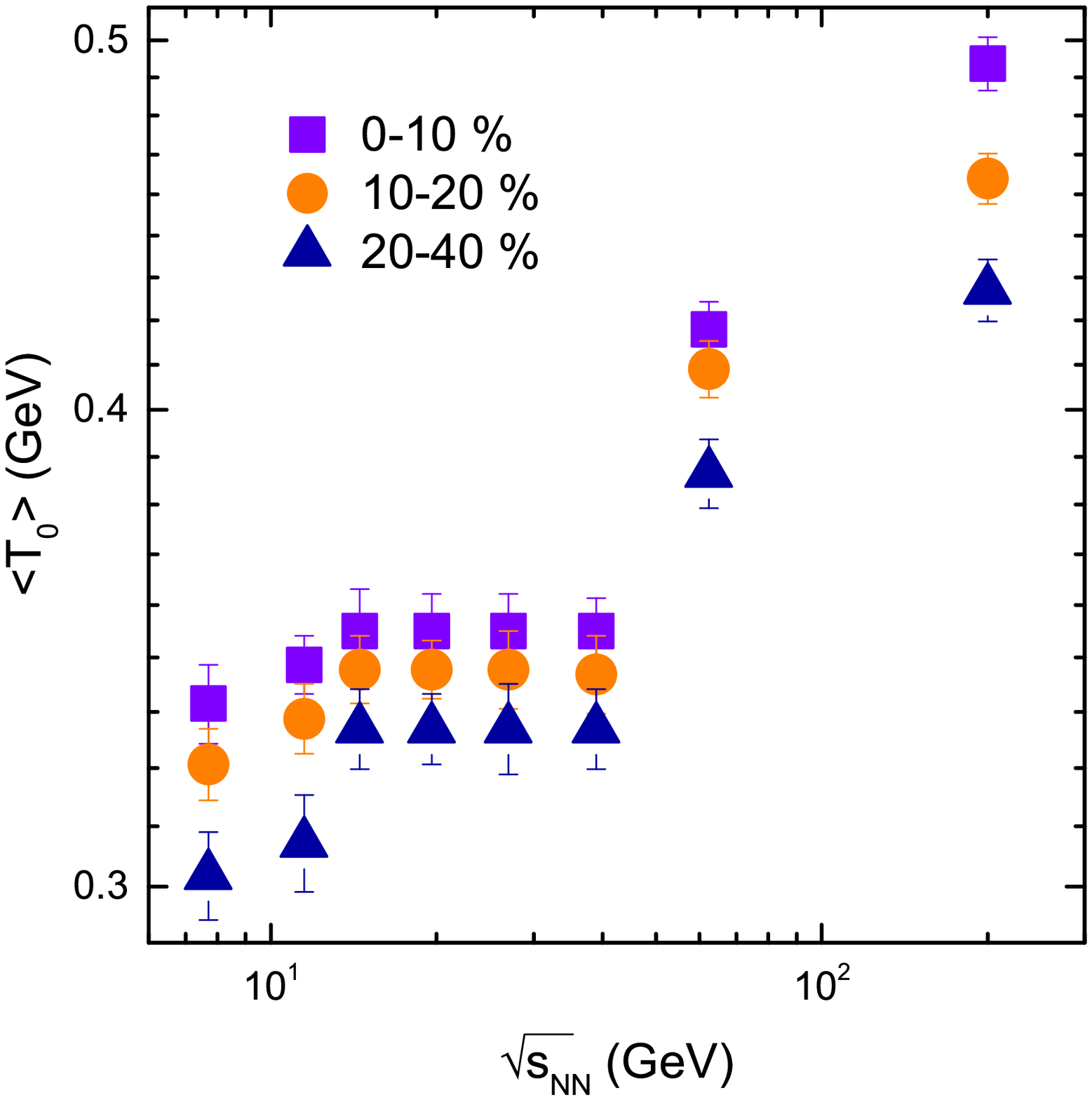}
\end{center}
Fig. 10. Dependence of average kinetic freeze-out temperature on
energy and centrality.
\end{figure*}

Before going to conclusion, we would like to point to out that we
obtained the initial temperature from root-mean-square $p_T$
taking analogy from string percolation model which is valid in the
present work because Hagedorn thermal model is a hybrid model. We
use the Monte-carlo simulation in the extraction of $T_i$. There
are many different methods by which we can extract $T_i$, such as
we can get $<p_T>$ and $\sqrt{<p^2_T>}$ from the combination of
data points also and fit function in the present work. Indeed, the
$p_T$ spectrum can be divided into two or three regions according
to the measured and un-measured $p_T$ ranges. One can use the data
points in the measured $p_T$ range and use the fit function only
to estimate to the un-measured $p_T$ range in order to obtain
$<p_T>$ and $\sqrt{<p^2_T>}$. The projectile and target
participant sources contribute equally to $<p_T>$ in each
nucleon-nucleon collision in AA collisions. According to the
framework of multi-source thermal model, [18, 66, 67] each
projectile and target source contribute a fraction of 1/2 to
$<p_T>$, i.e $<p_T>/2$, which contributes the thermal motion and
flow effect together. Let us consider $k_0(1-k_0)$ represents the
contribution fraction of thermal motion (flow effect), one can
define empirically
\begin{align}
T_0\equiv \frac{k_0<p_T>}{2},
\end{align}
and
\begin{align}
\beta_T \equiv \frac{(1-k_0)<p_T>}{2m_0\bar \gamma},
\end{align}
where $\bar \gamma$ represents the lorentz factor of the considered particle and

\begin{align}
k_0=0.30-0.01\ln (\sqrt{s_{NN}}),
\end{align}
is a parameterized representation. The effective temperature in a
recent work [68] is proportional to $<p_T>$ and the kinetic
freeze-out temperature is proportional to the effective
temperature, although the effective temperature in [68] and in our
present work is not the same. This method is diverting the focus
of the present work, so will not discuss it further.

In short, if we consider $O-xyz$ coordinate system, the momentum components in this system are
\begin{align}
p_x=p_T cos\theta,
\end{align}
\begin{align}
p_y=p_T sin\theta,
\end{align}
and
\begin{align}
p_x=p_T cot\theta=pcos\theta,
\end{align}
so, we have the relation between three components $p_x$, $p_y$ and $p_z$ of the momentum $p$ to be
\begin{align}
\sqrt{<p^2_x>}=\sqrt{<p^2_y>}=\sqrt{<p^2_z>}=\sqrt{<p^2_T>/2},
\end{align}
where $\sqrt{<p^2_x>}$, $\sqrt{<p^2_y>}$ and $\sqrt{<p^2_z>}$ are
root-mean square components and $T_i$ can be given by anyone of
these components. It should be noted that above isotropic
assumption is only performed in the source rest frame. However there are interactions among various sources,
and $p_z$ is lorentz boosted from one source to another one. This renders that there are anisotropic momentum
components in the final state.
 {\section{Conclusions}}
 The main observations and conclusions are summarized here.

(a) The transverse momentum spectra of the kaon, proton, deuteron
and triton produced in various centrality intervals in Au-Au
collisions at RHIC energies have been analyzed by Hagedorn thermal
model. The model results are in good agreement with the
experimental data in the special $p_T$ range measured by STAR
Collaboration.

(b) Effective temperature $T$, initial temperature $T_i$, kinetic
freeze-out temperature ($T_0$), and the mean transverse momentum
($<p_T>$) increases from 7.7 GeV to 14.5 GeV and then keeps
consistent up to 39 GeV, and once again grows at 62.4 and 200 GeV,
while the kinetic freeze-out volume ($V$) has an increasing trend
with increasing energy due to longer evolution time for partonic
system at higher energies.

(c) Due to low energy deposition, the system from 7.7 GeV to 14.5 GeV is
baryon influenced and there is no phase transition. The phase transition
in the system starts from 14.5 GeV in part volume and encounter from
baryon-influenced to meson-influenced due to the phase transition in
larger and larger volume in 14.5 GeV to 39 GeV. The identified boundaries
exists for the three energy ranges which exhibit three different mechanisms
of interaction and evolution process.

(d) The initial temperature and effective temperature for
triton are larger than deuteron, and the later is larger than
proton and kaon. Kaon has the lowest values for the temperatures
while the case for kinetic freeze-out volume is opposite where the
heaviest particle has the lowest $V$, which shows that the heavier
particles freeze-out early.

(e) The effective temperature ($T$), initial temperature ($T_i$),
kinetic freeze-out temperature ($T_0$), mean $<p_T>$ and kinetic
freeze-out volume ($V$) are observed to be larger in central
collisions and they decrease from central to periphery.

(f) The initial temperature is larger than the effective temperature, and the effective temperature larger than the kinetic freeze-out temperature.
The chemical freeze-out temperature ($T_{ch}$) lie between the initial and effective temperatures, and is nearly equal to the effective temperature and this sequence
agrees the order of time evolution of the interacting system, even though both the effective and kinetic freeze-out temperatures are extracted at the kinetic freeze-out.
\\
\\

{\bf Data availability}

The data used to support the findings of this study are included
within the article and are cited at relevant places within the
text as references.
\\
\\
{\bf Compliance with Ethical Standards}

The authors declare that they are in compliance with ethical
standards regarding the content of this paper.
\\
\\
{\bf Funding}

This research was funded by the National Natural Science Foundation of China grant
number 11875052, 11575190, and 11135011, and Ajman University Internal Research grant number
DGSR Ref. 2021-IRG-HBS-12, H.E.C Pakistan N.R.P.U Project Number 15785.
\\
\\
{\bf Acknowledgements}

We acknowledge the facilities provided by the University of Chinese Academy of Sciences China, Abdul Wali Khan
University Mardan, Ajman University Ajman U.A.E and Ghazi University D. G Khan for carrying out the research.
\\
\\

\end{multicols}
\end{document}